\PassOptionsToPackage{pdfpagelabels=false}{hyperref}
\documentclass[fleqn,usenatbib]{mnras}

\usepackage[T1]{fontenc}
\usepackage{ae,aecompl}
\usepackage{natbib}
\usepackage{graphicx}

\usepackage{txfonts}

\title[Polarisation of southern bright stars]{The linear polarisation of southern bright stars measured at the parts-per-million level}
\author[D.V.~Cotton et al.]{Daniel V. Cotton$^{1,2}$\thanks{E-mail: d.cotton@unsw.edu.au}, 
Jeremy Bailey$^{1,2}$, Lucyna Kedziora-Chudczer$^{1,2}$,
\newauthor Kimberly Bott$^{1,2}$, P.W.~Lucas$^3$, J.H.~Hough$^3$ and Jonathan P. Marshall$^{1,2}$ \\
$^{1}$School of Physics, UNSW Australia, NSW 2052, Australia.\\
$^{2}$Australian Centre for Astrobiology, UNSW Australia, NSW 2052, Australia.\\
$^{3}$Centre for Astrophysics Research, Science \& Technology Research Institute, University of Hertfordshire, Hatfield, AL10 9AB, UK.}
\begin{document}

\date{Accepted . Received ; in original form }

\pagerange{\pageref{firstpage}--\pageref{lastpage}} \pubyear{2015}

\maketitle

\label{firstpage}

\begin{abstract}We report observations of the linear polarisation of a sample of 50 nearby southern bright stars measured to a median sensitivity of $\sim$4.4 $\times 10^{-6}$. We find larger polarisations and more highly polarised stars than in the previous PlanetPol survey of northern bright stars. This is attributed to a dustier interstellar medium in the mid-plane of the Galaxy, together with a population containing more B-type stars leading to more intrinsically polarised stars, as well as using a wavelength more sensitive to intrinsic polarisation in late-type giants. Significant polarisation had been identified for only six stars in the survey group previously, whereas we are now able to deduce intrinsic polarigenic mechanisms for more than twenty.

The four most highly polarised stars in the sample are the four classical Be stars ($\alpha$ Eri, $\alpha$ Col, $\eta$ Cen and $\alpha$ Ara). For the three of these objects resolved by interferometry, the position angles are consistent with the orientation of the circumstellar disc determined. We find significant intrinsic polarisation in most B stars in the sample; amongst these are a number of close binaries and an unusual binary debris disk system. However these circumstances do not account for the high polarisations of all the B stars in the sample and other polarigenic mechanisms are explored. Intrinsic polarisation is also apparent in several late type giants which can be attributed to either close, hot circumstellar dust or bright spots in the photosphere of these stars. Aside from a handful of notable debris disk systems, the majority of A to K type stars show polarisation levels consistent with interstellar polarisation. \end{abstract}


\section{Introduction}

The measured linear polarisation of starlight falls into two categories: it is either intrinsic to the star and its immediate environment (i.e. the star system), or it is interstellar in origin. Interstellar polarisation is the result of interstellar dust particles aligning with the Galactic magnetic field. Studies of this polarisation reveal details of the dust distribution and the Galactic magnetic field structure \citep{heiles96} as well as the size and nature of the dust particles \citep{whittet92,kim94}. Linear polarimetry is therefore a powerful technique for investigating the interstellar medium.

Linear polarisation measurements of stars have, until fairly recently, been limited to sensitivities in fractional polarisation of $\sim$10$^{-4}$. At this level most nearby ($d$ $<$ 100 pc) stars are found to be unpolarised \citep{tinbergen82,leroy93a,leroy93b,leroy99}. A new generation of high precision polarimeters has recently been developed that allow polarisation measurements of starlight at the parts per million level \citep{hough06,wiktorowicz08,bailey15,wiktorowicz15}. With such instruments it is possible to detect significant interstellar polarisation even in nearby stars. \cite{bailey10} reported such a survey of 49 bright nearby northern hemisphere stars (the PlanetPol survey) and found that the majority of stars show significant polarisation, in most cases consistent with an interstellar origin. 

In this paper we present the first survey of the polarisation of southern bright nearby stars measured at the parts per million level. The survey utilised the newly commissioned HIPPI (HIgh Precision Polarimetric Instrument) \citep{bailey15} to observe 50 of the brightest southern stars within 100 pc. The observing methods are described in detail in Section \ref{sec_survey}. In contrast to the PlanetPol survey \citep{bailey10} described above, the southern sample reported here contains many stars that are intrinsically polarised.

Intrinsic polarisation is well known to be present in Be stars as a result of scattering from a transient gaseous circumstellar disc and our sample includes four classical Be stars. These are indeed the four highest polarisations measured. However, compared to other stars we observed we also find higher than average polarisations in several B stars, not known to be Be stars. These results are discussed in Sections \ref{sec_be_stars} and \ref{sec_b_stars} respectively. Other sources of intrinsic polarisation found in the sample are the presence of a debris disk\footnote{Throughout this paper disk is used in reference to debris disks as preferred by the debris disk research community, disc being used otherwise.} around the star (Section \ref{sec_debris_disk}), and polarisation, probably mostly due to dust scattering, in late type giants (Section \ref{sec_giants}).

Another area of interest is the level to which normal stars are intrinsically polarised. The quiet Sun has been directly measured, and shows polarisation of $<$3 $\times 10^{-7}$ \citep{kemp87}. Starspots in active stars will show higher polarisation \citep{strassmeier09,wiktorowicz09}. As well as being interesting in their own right, starspots can complicate the growing and exciting field of exoplanet polarimetry \citep{wiktorowicz09}. 

Polarimetry can be used to detect unresolved hot-Jupiter type planets \citep{seager00,lucas06,lucas09}, as a differential technique to detect planets in imaging \citep{schmid05,keller06} or transit observations \citep{kostogryz15}, and to characterise exoplanet atmospheres \citep{bailey07,kopparla14}. In each of these applications the signal from the planet is small. The effectiveness of the technique will depend upon an understanding of non-planetary sources of polarisation as much as planetary ones. For this reason an investigation and characterisation of the primary sources of stellar linear polarisation is important.

\section{Observations}
\label{sec_survey}

\subsection{The sample stars}
\label{sec_sample}

Stars selected for observation were south of the equator, were at a distance of less than 100 pc\footnote{One star, HIP 89931, has a distance greater than 100 pc as a result of revision of its parallax after the sample was selected.}, and have V magnitude brighter than 3.0\footnote{A V magnitude of 3.0 corresponds to an absolute magnitude of -2.0 at 100 pc.}. 54 stars met these criteria and we have observed 50 of them. The stars observed are listed in Table \ref{tab_stars}. This table gives the V magnitude and spectral type as listed in the SIMBAD database\footnote{HIP 60718, has no listed luminosity class in SIMBAD, and so instead we use the spectral type given by \citet{bartkevicius01}. There are a number of multiple systems listed, HIP 93506 and HIP 65474, for example, are both noteworthy double star systems where both stars are of B spectral type, where the combined spectral type is given by SIMBAD this is what is given in Table \ref{tab_stars}. If a companion also falls within the HIPPI aperture this is marked with an '*' in the succeeding column.}, the distance derived from the Hipparcos catalogue parallax \citep{perryman97,vanleeuwen07}, the Galactic coordinates (also from the Hipparcos catalogue), and rotational velocity ($V$~sin~$i$) from a variety of sources as indicated in Table \ref{tab_stars}. Where a distinction has been made in the reference, the value given is for the primary in multiple systems. The number of companions that fall within the aperture was determined in the first instance from Burnham's Celestial Handbook \citep{burnham78a,burnham78c,burnham78b} and from \citet{eggleton08}, then followed up with a variety of sources as indicated in the footnote of Table \ref{tab_stars}.

\begin{table*}
\label{tab_stars}
\caption{Properties of survey stars.}
\tabcolsep 2.5 pt
\centering
\begin{tabular}{rrlrlcrccrrll}
\hline
HIP  & BS  &Other Names  & V\hspace{1.5 mm} & Spectral & Comp & Dist & RA &  Dec & \multicolumn{2}{c}{Galactic} & \hspace{4.5 mm}$V$~sin~$i$ & $F_{\rm IR}/F_{\star}^b$\\
     &     &           &  mag   &  Type & in Ap$^a$ & (pc) & (hh mm) & (dd mm) &   Long & Lat  &  \hspace{4.5 mm}(km/s) & ($\times10^{-6}$)\\
     \hline                
2021 & 98  &  $\beta$ Hyi               & 2.79  &  G0V     & & 7.5  &  00 26 &  $-$77 15 &  304.77 & $-$39.78 & \hspace{3 mm}4.1 $\pm$ \hspace{1.5 mm}1.0$^{dM}$  &  \\
2081 & 99  &  $\alpha$ Phe              & 2.37  &  K0.5IIIb & & 25.9 & 00 26 &  $-$42 18 &  320.00 & $-$73.98 & \hspace{7.5 mm}$<$ \hspace{1.5 mm}1.2$^{dM}$  &       \\
3419 & 188 &  $\beta$ Cet               & 2.01  &  K0III   & & 29.5 &  00 44 &  $-$17 59 &  111.33 & $-$80.68 & \hspace{3 mm}3.6 $\pm$ \hspace{1.5 mm}1.0$^{dM}$  &  \\
7588 & 472 &  $\alpha$ Eri, Achernar    & 0.46  & B6Vep    & & 42.9 &  01 38 &  $-$57 14 &  290.84 & $-$58.79 & 207 \hspace{2.3 mm}$\pm$ \hspace{1.5 mm}9$^{Y}$     & \hspace{3 mm}0.027\\
9236 & 591 &  $\alpha$ Hyi              & 2.84  &  F0IV    & & 22.0 &  01 59 &  $-$61 34 &  289.44 & $-$53.76 & 118 \hspace{2.3 mm}$\pm$ 12$^{ZR}$                  & \hspace{1.5 mm}34.24\\
13847 & 897 & $\theta$ Eri              & 2.90  &  A4III   & & 49.4 &  02 58 &  $-$40 18 &  247.86 & $-$60.74 & \hspace{1.5 mm}70 \hspace{2.3 mm}$\pm$ 17$^{R}$ &  \\
18543 & 1231 & $\gamma$ Eri             & 2.94  &  M1IIIb  & & 62.3 &  03 58 &  $-$13 30 &  205.16 & $-$44.47 & \hspace{3 mm}3.8 $\pm$ \hspace{1.5 mm}0.3$^{M}$   & \\
26634 & 1956 & $\alpha$ Col             & 2.65  &  B9Ve    & & 80.1 &  05 40 &  $-$34 04 &  238.81 & $-$28.86 & 184 \hspace{2.3 mm}$\pm$ \hspace{1.5 mm}5$^{Y}$     & \hspace{3 mm}1.484\\
30438 & 2326 & $\alpha$ Car, Canopus    & $-$0.74 & A9II   & & 94.8 &  06 24 &  $-$52 42 &  261.21 & $-$25.29 & \hspace{3 mm}9.0 $\pm$ \hspace{1.5 mm}1.0$^{R}$   & \hspace{3 mm}0.033\\
32349 & 2491 & $\alpha$ CMa, Sirius     & $-$1.46 & A1V    & & 2.6  &  06 45 &  $-$16 43 &  227.23 & $-$08.89 & \hspace{1.5 mm}16 \hspace{2.3 mm}$\pm$ \hspace{1.5 mm}1$^{R}$ &   \\
39757 & 3185 & $\rho$ Pup               & 2.81 &   F5II    & & 19.5 &  08 08 &  $-$14 18 &  243.15 & 04.40    & \hspace{1.5 mm}15 \hspace{2.3 mm}$\pm$ \hspace{1.5 mm}1$^{GG}$ & 160.7\\
42913 & 3485 & $\delta$ Vel             & 1.95 &   A1Va(n) &*& 24.7 &  08 45 &  $-$54 43 &  272.08 & $-$07.37 & 150 \hspace{2.3 mm}$\pm$ 15$^{ZR}$ & \hspace{3 mm}4.567 \\
45238 & 3685 & $\beta$ Car              & 1.69 &   A1III   & & 34.7 &  09 13 &  $-$69 43 &  285.98 & $-$14.41 & 146 \hspace{2.3 mm}$\pm$ \hspace{1.5 mm}2$^{D}$     & \hspace{3 mm}6.700\\
46390 & 3748 & $\alpha$ Hya, Alphard    & 1.97 &   K3II-III & & 54.0 & 09 28 &  $-$08 40 &  241.49 & 29.05    & \hspace{7.5 mm}$<$ \hspace{1.5 mm}1.0$^{dM}$  & \hspace{3 mm}3.359\\
52727 & 4216 & $\mu$ Vel                & 2.69 &   G6III   &*& 35.9 &  10 47 &  $-$49 25 &  283.03 & 08.57    & \hspace{3 mm}6.4 $\pm$ \hspace{1.5 mm}1.0$^{dMM}$ & \hspace{1.5 mm}19.94\\
59803 & 4662 & $\gamma$ Crv             & 2.58 &   B8III   & & 47.1 &  12 16 &  $-$17 33 &  290.98 & 44.50    & \hspace{1.5 mm}30 \hspace{2.3 mm}$\pm$ \hspace{1.5 mm}9$^{A}$ &   \\
60718 &      & $\alpha$ Cru             & 0.81 &   B0.5IV &$*^c$& 98.7 &  12 27 &  $-$63 06 &  300.13 & $-$00.36 & \hspace{1.5 mm}88 \hspace{2.3 mm}$\pm$ \hspace{1.5 mm}5$^{GG}$ & \hspace{3 mm}0.001 \\
60965 & 4757 & $\delta$ Crv             & 2.94 &   A0IV(n) & & 26.6 &  12 30 &  $-$16 31 &  295.47 & 46.04    & 236 \hspace{2.3 mm}$\pm$ 24$^{ZR}$                  & \\
61084 & 4763 & $\gamma$ Cru             & 1.64 &   M3.5III & & 27.2 &  12 31 &  $-$57 07 &  300.17 & 05.65    & \hspace{3 mm}4.7 $\pm$ \hspace{1.5 mm}0.9$^{C}$   & \\
61359 & 4786 & $\beta$ Crv              & 2.64 &   G5II    & & 44.7 &  12 34 &  $-$23 24 &  297.87 & 39.31    & \hspace{3 mm}4.2 $\pm$ \hspace{1.5 mm}1.0$^{dMM}$ & \hspace{3 mm}5.154 \\
61585 & 4798 & $\alpha$ Mus             & 2.68 &   B2IV-V  & & 96.7 &  12 37 &  $-$69 08 &  301.66 & $-$06.30 & 114 \hspace{2.3 mm}$\pm$ 11$^{W}$                   & \hspace{3 mm}0.179 \\
61932 & 4819 & $\gamma$ Cen             & 2.17 &   A1IV+A0IV &*& 39.9 &  12 42 &  $-$48 58 &  301.25 & 13.88    & \hspace{1.5 mm}79 \hspace{2.3 mm}$\pm$ \hspace{1.5 mm}3$^{GG}$ &\\
61941 &      & $\gamma$ Vir             & 2.74 &   F0IV+F0IV &*& 11.7 & 12 42 & $-$01 27 &  297.83 & 61.33    & \hspace{1.5 mm}36 \hspace{2.3 mm}$\pm$ \hspace{1.5 mm}4$^{ZR}$ &\\
64962 & 5020 & $\gamma$ Hya             & 3.00 &   G8III   & & 41.0 &  13 19 &  $-$23 10 &  311.10 & 39.26    & \hspace{3 mm}3.7 $\pm$ \hspace{1.5 mm}1.0$^{dM}$  & \hspace{3 mm}2.931\\
65109 & 5028 & $\iota$ Cen              & 2.73 &   A2V     & & 18.0 &  13 20 &  $-$36 43 &  309.41 & 25.79    & \hspace{1.5 mm}90 \hspace{2.3 mm}$\pm$ \hspace{1.5 mm}2$^{D}$ & \hspace{1.5 mm}14\\
65474 & 5056 & $\alpha$ Vir, Spica      & 0.97 &   B1III-IV &*& 77.0 & 13 25 &  $-$11 10 &  316.11 & 50.84    & 140 \hspace{2.3 mm}$\pm$ \hspace{1.5 mm}6$^{GG}$    & \hspace{3 mm}0.003\\
68933 & 5288 & $\theta$ Cen             & 2.05 &   K0III   & & 18.0 & 14 07 &   $-$36 22 &  319.45 & 24.08    & \hspace{3 mm}1.2$^{GG}$ &  \\
71352 & 5440 & $\eta$ Cen               & 2.31 &   B1.5Vne & & 93.7 &  14 36 &  $-$42 09 &  322.77 & 16.67    & 291 \hspace{2.3 mm}$\pm$ 32$^{Y}$                   & \hspace{3 mm}0.443\\
71683 & 5459 & $\alpha$ Cen             & 0.10 &   G2V+K1V &*& 1.3 &   14 40 &  $-$60 50 &  315.73 & $-$00.68 & \hspace{3 mm}2.3 $\pm$ \hspace{1.5 mm}0.5$^{We}$  & \\
72622 & 5531 & $\alpha^2$ Lib           & 2.75 &   A3V     &*& 23.2 &  14 50 &  $-$16 02 &  340.32 & 38.01    & 102 \hspace{2.3 mm}$\pm$ 10$^{ZR}$                  & \hspace{1.5 mm}16.25\\
74785 & 5685 & $\beta$ Lib              & 2.62 &   B8Vn    &*& 56.8 &  15 17 &  $-$09 22 &  352.02 & 39.23    & 250 \hspace{2.3 mm}$\pm$ 24$^{A}$                   & \hspace{3 mm}3.652\\
74946 & 5671 & $\gamma$ TrA             & 2.89 &   A1V     & & 56.4 &  15 19 &  $-$68 41 &  315.71 & $-$09.55 & 199 \hspace{2.3 mm}$\pm$ 20$^{ZR}$                  & \hspace{3 mm}5.391\\
77952 & 5897 & $\beta$ TrA              & 2.85 &   F1V     & & 12.4 &  15 55 &  $-$63 26 &  321.85 & $-$07.52 & \hspace{1.5 mm}92$^{Ma}$          & \\
79593 & 6056 & $\delta$ Oph             & 2.75 &   M0.5III & & 52.5 &  15 14 &  $-$03 41 &  8.84   & 32.20    & \hspace{3 mm}7.0 $\pm$ \hspace{1.5 mm}0.0$^{M}$  & \\
82396 & 6241 & $\epsilon$ Sco           & 2.29 &   K1III   & & 19.5 &  16 50 &  $-$34 18 &  348.81 & 06.56    & \hspace{7.5 mm}$<$ \hspace{1.5 mm}1.0$^{dM}$  &       \\
84012 & 6378 & $\eta$ Oph               & 2.42 &   A2IV-V  &*& 27.1 &  17 10 &  $-$15 44 &  6.72   & 14.01    & \hspace{1.5 mm}23 \hspace{2.3 mm}$\pm$ \hspace{1.5 mm}2$^{ZR}$  & \hspace{1.5 mm}19.01\\
85792 & 6510 & $\alpha$ Ara             & 2.95 &   B2Vne   & & 82.0 &  17 32 &  $-$49 52 &  340.75 & $-$08.83 & 269 \hspace{2.3 mm}$\pm$ 13$^{Y}$                   & \hspace{1.5 mm}15.62\\
86228 & 6553 & $\theta$ Sco             & 1.86 &   F1III   & & 92.1 &  17 37 &  $-$43 00 &  347.14 & $-$05.98 & 125$^{vB}$ & \hspace{3 mm}3.023      \\
88635 & 6746 & $\gamma^2$ Sgr           & 2.99 &   K1III   & & 29.7 &  18 06 &  $-$30 25 &  0.92   & $-$04.54 & \hspace{7.5 mm}$<$ \hspace{1.5 mm}1.0$^{dM}$   & 146.7      \\
89931 & 6859 & $\delta$ Sgr             & 2.71 &   K3IIIa  & & 106.6 & 18 21 &  $-$29 50 &  3.00   & $-$70.15 & \hspace{3 mm}3.6 $\pm$ \hspace{1.5 mm}1.0$^{dM}$   & \hspace{3 mm}1.305\\
90185 & 6879 & $\epsilon$ Sgr           & 1.85 &   B9.5III &*& 43.9 &  18 24 &  $-$34 23 &  359.19 & $-$09.81 & 175$^{vB}$ &  \hspace{1.5 mm}31 \\
90496 & 6913 & $\lambda$ Sgr            & 2.81 &   K0IV    & & 24.0 &  18 28 &  $-$25 25 &  7.66   & $-$06.52 & \hspace{7.5 mm}$<$ \hspace{1.5 mm}1.1$^{dM}$   &   \\
92855 & 7121 & $\sigma$ Sgr             & 2.06 &   B2V     & & 69.8 &  18 55 &  $-$26 18 &  9.56   & $-$12.43 & 205$^{vB}$ & \hspace{3 mm}0.025 \\
93506 & 7194 & $\zeta$ Sgr              & 2.61 &   A2.5Va  &*& 27.0 &   19 03 &  $-$29 53 &  6.84   & $-$15.35 & \hspace{1.5 mm}77 \hspace{2.3 mm}$\pm$ \hspace{1.5 mm}8$^{ZR}$  & \hspace{3 mm}7.111\\
100751 & 7790 & $\alpha$ Pav            & 1.91 &   B2IV    &*& 54.8 &  20 26 &  $-$56 44 &  340.9  & $-$35.19 & \hspace{1.5 mm}15 \hspace{2.3 mm}$\pm$ \hspace{1.5 mm}3$^{GG}$  & \hspace{3 mm}0.038\\
107556 & 8322 & $\delta$ Cap            & 2.83 &   A7III   &*& 11.9 &  21 47 &  $-$16 08 &  37.60  & $-$46.01 & \hspace{1.5 mm}93 \hspace{2.3 mm}$\pm$ 10$^{GG}$   & \\
109268 & 8425 & $\alpha$ Gru            & 1.71 &   B6V     & & 31.0 &  22 08 &  $-$46 58 &  349.99 & $-$52.47 & 259 \hspace{2.3 mm}$\pm$ 26$^{ZR}$  & \\
110130 & 8502 & $\alpha$ Tuc            & 2.82 &   K3III   &*& 61.3 &  22 19 &  $-$60 16 &  330.22 & $-$47.96 & \hspace{3 mm}1.9 $\pm$ \hspace{1.5 mm}1.3$^{dM}$  & \hspace{3 mm}6.579 \\
112122 & 8636 & $\beta$ Gru             & 2.11 &   M5III   & & 54.3 &  22 43 &  $-$46 53 &  346.27 & $-$57.95 & \hspace{3 mm}2.5 $\pm$ \hspace{1.5 mm}1.5$^{Z}$  & \\
113368 & 8728 & $\alpha$ PsA, Fomalhaut & 1.16 &   A4V     & & 7.7  &  22 58   &  $-$29 37 &  20.49  & $-$64.91 & \hspace{1.5 mm}93 \hspace{2.3 mm}$\pm$ \hspace{1.5 mm}9$^{ZR}$ & \hspace{1.5 mm}22\\
\hline
\end{tabular}
\begin{flushleft}
a - An asterisks indicates a companion also lies within the aperture. Sources for these determinations are: \cite{burnham78a,burnham78c,burnham78b}, \cite{eggleton08}, \cite{pourbaix04}; Spica - \cite{harrington09}; $\epsilon$ Sgr - \cite{hurbig01}; $\beta$ Lib - \cite{roberts07}.  \\
b - Fractional infrared excess from \cite{mcdonald12}, except for $\iota$ Cen and Fomalhaut \citep{marshall15}, and $\epsilon$ Sgr for which we derived the fractional luminosity based on excesses reported by \citet{rhee07} and \citet{chen14}. \\
c - There are at least two stars within the aperture, there maybe as many as four (see Section \ref{sec_alpha_cru}). \\
Rotational velocity references: dM - \cite{demedeiros14}, Y - \cite{yudin01}, ZR - \cite{zorec12}, R - \cite{royer02}, M - \cite{massarotti08}, dMM - \cite{demedeiros99}, A - \cite{abt02}, GG - \cite{glebocki05}, C - \cite{cummings98}, W - \cite{wolff07}, D - \cite{diaz11}, We - \cite{weise10}, Ma - \cite{mallik03}, vB - \cite{vanbelle12}, Z - \cite{zamarov08}. \\
\end{flushleft}
\end{table*}

Previous polarisation measurements for the stars are listed in Table \ref{tab_previous} and have been taken from the agglomerated polarisation catalogue of \citet{heiles00}. Only the degree of polarisation is listed. The position angle can be found in the original catalogue, but for almost all the measurements the polarisation is not significant and the position angle is therefore meaningless. The only stars with significant polarisations in this catalogue are the four Be stars, which will be discussed later, the SRB (semi-regular) type pulsating variable HIP 112122 ($\beta$ Gru) which has a listed polarisation angle of 155.2$\degr$$\pm$4.6, and possibly the short period binary and variable of the $\beta$ Cep type HIP 65474 (Spica) with a listed polarisation angle of 147.0$\degr$$\pm$0.0.

The \citet{heiles00} catalogue does not include the slightly more accurate previous polarisation measurements of nearby stars made by \citet{tinbergen82}. This study includes a number of the stars in our sample and is therefore listed separately in Table \ref{tab_previous}. The measurements of Tinbergen were made in three colour bands. We have tabulated the averaged band I and II measurements. Of the stars observed by \citet{tinbergen82} only HIP 89931 ($\delta$ Sgr) and Spica have a significant polarisation.

\begin{table}
\caption{Previous polarisation measurements.}
\centering
\begin{tabular}{llrrrr}
\hline
HIP & Polarisation (Heiles) & \multicolumn{2}{c}{Polarisation (Tinbergen)$^a$} \\
    &  $p$ (\%)       &   $q$ (ppm)   &   $u$ (ppm)       \\
\hline
2021  &   0.050$\pm$0.035   &     60 $\pm$ 60    &        \hspace{1.5 mm}0 $\pm$ 60      \\
2081  &   0.011$\pm$0.006   &     60 $\pm$ 60    &     $-$70 $\pm$ 60      \\
3419  &   0.004$\pm$0.005   &     40 $\pm$ 60    &     $-$60 $\pm$ 60      \\
7588$^{b}$ & 0.011$\pm$0.002 &     10 $\pm$ 60    &     $-$10 $\pm$ 60      \\
9236  &                     &     30 $\pm$ 60    &     $-$10 $\pm$ 60      \\
13847 &   0.003$\pm$0.003   &                  &            \\
26634$^{b}$ & 0.150$\pm$0.100 &                 &             \\
32439 &                     &     \hspace{1.5 mm}0 $\pm$ 60    &     $-$40 $\pm$ 60      \\
46390 &   0.150$\pm$0.120   &     30 $\pm$ 60    &        10 $\pm$ 60      \\
52727 &                     &     30 $\pm$ 60    &     $-$40 $\pm$ 60      \\
59803 &   0.050$\pm$0.120   &                  &            \\
60965 &   0.030$\pm$0.120   &                  &                       \\
61359 &                     &    120 $\pm$ 60    &     $-$30 $\pm$ 60      \\
61941 &   0.030$\pm$0.120   &     60 $\pm$ 60    &        70 $\pm$ 60      \\
65109 &                     &  $-$30 $\pm$ 60    &       110 $\pm$ 60      \\
65474$^c$ & 0.010$\pm$0.000 & $-$220 $\pm$ 60    &     $-$10 $\pm$ 60    \\
      &                     & $-$150 $\pm$ 60    &     $-$60 $\pm$ 60    \\
71352$^{b}$ & 0.040$\pm$0.100 &                 &                  \\
71683 &                     &  $-$70 $\pm$ 60    &        30 $\pm$ 60      \\
74785 &   0.050$\pm$0.120   &                  &             \\
77952 &                     &  $-$40 $\pm$ 60    &        40 $\pm$ 60      \\
79593 &   0.020$\pm$0.120   &                  &             \\
82396 &                     &     60 $\pm$ 60    &     $-$90 $\pm$ 60      \\
84012 &   0.026$\pm$0.004   &                  &             \\
85792$^{b}$ & 0.610$\pm$0.035 &                 &           \\
86228 &   0.030$\pm$0.035   &                  &             \\
89931 &                     & $-$110 $\pm$ 60    &    $-$520 $\pm$ 60      \\
90496 &                     &     90 $\pm$ 60    &    $-$150 $\pm$ 60      \\
107556 &                    &  $-$40 $\pm$ 60    &     $-$40 $\pm$ 60      \\
109268 &  0.003$\pm$0.005   &  $-$70 $\pm$ 60    &        10 $\pm$ 60      \\
110130 &  0.011$\pm$0.004   &                  &               \\
112122 &  0.031$\pm$0.005   &                  &             \\
113368 &  0.005$\pm$0.005   &     10 $\pm$ 60    &    $-1$10 $\pm$ 60      \\
\hline
\end{tabular}
\begin{flushleft}
a - The polarisations were originally reported with units of 10$^{-5}$. \\
b - Classical Be star. \\
c - Spica appears in Tinbergen's table twice; observed from Hartebeesportdam (top) and LaSilla (bottom). \\
\end{flushleft}
\label{tab_previous}
\end{table}
     
In addition to the measurement listed in Table \ref{tab_previous}, it might be noted that Sirius (HIP 32349) also has a reported measurement in \citet{bailey15} where it was used as a low polarisation standard to determine telescope polarisation. When the telescope polarisation is subtracted the polarisation of Sirius is 3$\pm$4 ppm.

\subsection{Observation methods}

The observations were obtained with the HIPPI (High Precision Polarimetric Instrument) polarimeter \citep{bailey15} over the course of four observing runs (May 8th to May 12th 2014, August 28th to September 2nd 2014, May 21st to 26th 2015 and June 25th to 29th 2015) at the 3.9 m Anglo Australian Telescope (AAT). The AAT is located at Siding Spring Observatory near Coonabarabran in New South Wales, Australia. HIPPI was mounted at the f/8 Cassegrain focus of the telescope where it had an aperture size of 6.7$\arcsec$. The seeing varied between $\sim$1.5$\arcsec$ and $>$ 4$\arcsec$ with a typical seeing around 2-2.5$\arcsec$. 

HIPPI is a high precision polarimeter, with a sensitivity in fractional polarisation of $\sim$4.3 $\times 10^{-6}$ on stars of low polarisation and a precision of better than 0.01\% on highly polarised stars \citep{bailey15}. It achieves this by the use of a Ferroelectric Liquid Crystal (FLC) modulator operating at a frequency of 500 Hz to eliminate the effects of variability in the atmosphere. From the modulator the light passes through a Wollaston prism that acts as a beam splitter, directing the light into two Photo Multiplier Tube (PMT) detectors. Second stage chopping, to reduce systematic effects, is accomplished by rotating the entire back half of the instrument after the filter wheel through 90$\degr$ in an ABBA pattern, with a frequency of once per 40 seconds. An observation of this type measures only one Stokes parameter of linear polarisation. To get the orthogonal Stokes parameter the entire instrument is rotated through 45$\degr$ and the sequence repeated. The rotation is performed using the AAT's Cassegrain instrument rotator. In practice we also repeat the observations at telescope position angles of 90$\degr$ and 135$\degr$ to allow removal of instrumental polarisation. The stars in the survey were observed for a total integration time of 12 minutes in each ($Q$ and $U$) Stokes parameter per observation. This gave a median sensitivity in fractional polarisation of $\sim$4.4 $\times 10^{-6}$. The observing, calibration and data reduction methods are described in full detail in \citet{bailey15}. 

Typically, a sky measurement was acquired at each telescope position angle an object was observed in, and subtracted from the measurement. The duration of the sky measurements was 3 minutes per Stokes parameter. For particularly bright objects (or those known to be highly polarised) observed under a moonless sky, the sky signal is negligible and a dark measurement was sufficient for calibration purposes. Dark measurements are obtained with the use of a blank in the filter wheel and the instrument otherwise configured identically. Whether a sky subtraction or only a dark subtraction was carried out is indicated in Table \ref{tab_results}. These subtractions were carried out as the first step of the data reduction routine. 

During the observations at the AAT, zero point calibration was carried out by reference to the average of a set of observed stars that have been measured as having negligible polarisation. This method was used owing to a lack of other available unpolarised standard stars in the southern hemisphere, and allowed us to determine the telescope polarisation for the equatorially mounted AAT. Using this method the  polarisation of the telescope was determined as 48$\pm$5 $\times 10^{-6}$ during the Aug-Sep run \citep{bailey15}. This value was used for correction of both 2014 runs, with the uncertainty in the measurement contributing to the stated errors. Shortly before the May 2015 run the telescope mirror was re-aluminised necessitating a new determination of the telescope polarisation. This procedure was carried out as before, with the details given in Table \ref{tab_tp}. HIPPI was left in place at the AAT Cassegrain focus between the May and June 2015 runs. Three additional measurements of low polarisation standards ($\beta$ Leo, and two of BS 5854) were made in June. To increase the precision of the determination for June 2015 the average of all the May and June measurements was adopted as the telescope polarisation for that run -- this is reasonable given the negligible difference in the determinations. The telescope polarisation was found to be 36.5$\pm$1.2 $\times 10^{-6}$.

\begin{table}
\caption{Low polarisation star measurements to determine telescope polarisation (TP) for the May and June 2015 runs in the $g^{\rm \prime}$ filter. The adopted TP for June 2015 uses all seven measurements, that for May 2015 just those acquired in that month.}
\centering
\begin{tabular}{llcll}
\hline
Star &  Date &     $p$ (ppm) &   \hspace{2 mm}$\theta$ ($\degr$) \\
\hline
Sirius  &    23 May &   39.2 $\pm$ 1.4 & 91.0 $\pm$ 2.1    \\
        &    24 May &   40.2 $\pm$ 0.8 & 89.5 $\pm$ 1.1    \\
BS 5854 &    22 May &   30.4 $\pm$ 5.1 & 70.2 $\pm$ 9.7    \\
        &    26 May &   37.6 $\pm$ 4.0 & 84.4 $\pm$ 6.1    \\
\hline
Adopted TP & May 2015 & 35.5 $\pm$ 1.4 & 84.7 $\pm$ 2.3    \\
\hline
BS 5854 &   26 Jun  &   43.3 $\pm$ 4.4 & 93.4 $\pm$ 5.9    \\ 
        &   27 Jun  &   32.9 $\pm$ 3.9 & 70.6 $\pm$ 6.8    \\ 
$\beta$ Leo& 26 Jun &   43.6 $\pm$ 2.4 & 87.0 $\pm$ 3.2    \\
\hline
Adopted TP & Jun 2015 & 36.5 $\pm$ 1.2 & 84.8 $\pm$ 1.4    \\
\hline
\end{tabular}
\label{tab_tp}
\end{table} 

A number of stars with known high polarisations ($\sim$1-5\%) were observed during the Aug-Sep run, and used to help calibrate the modulation efficiency of the telescope as reported by \citet{bailey15}. The same stars were used to determine the position angle zero-point for the Aug-Sep run, whereas a different set were needed for the other runs; these were HD 147084, HD 154445, HD 160529 and HD 187929 in May 2014; HD 147084 and HD 154445 in May 2015; and HD 147084 and HD 154445 in June 2015. The precision of each determination is less than 1 degree, based on the consistency of the calibration provided by the different reference stars which themselves have uncertainties of this order.

HIPPI has been tuned to be most sensitive at the blue end of the visible spectrum, in order to investigate polarisation induced in exoplanet atmospheres by Rayleigh scattering from clouds \citep{bailey15}. This it achieves through the use of PMTs with ultra-bialkali photocathodes, giving a quantum efficiency of 43\% at 400 nm. HIPPI is equipped with a number of filters. The initial observations reported here were carried out exclusively using the SDSS $g^{\rm \prime}$ (400-550 nm) filter. 

The $g^{\rm \prime}$ band is centred on 475 nm and is 150 nm in width, which results in the precise effective wavelength\footnote{Effective wavelength is the intensity weighted mean of the wavelength of observed photons.} and modulation efficiency changing with star colour. Table \ref{tab_eff} gives the effective wavelength and modulation efficiency for various spectral types based on a bandpass model as described in \citet{bailey15}. It combines the attenuation of the atmosphere, as well as PMT response and other instrumental responses as well as that of the $g^{\rm \prime}$ filter. No interstellar extinction (i.e. $E(B-V)=0$) has been applied here. This is reasonable as at 100 pc the extinction is well below that which can actually be measured by photometry \citep{bailey10}. The efficiencies in Table \ref{tab_eff} were applied to the raw $Q/I$ and $U/I$ polarisations of each object according to its spectral type given in Table \ref{tab_stars} (we apply a linear interpolation between the given types).

\begin{table}
\caption{Effective wavelength and modulation efficiency for different spectral types according to bandpass model.}
\centering
\begin{tabular}{lcc}
\hline
Spectral    &   Effective       &   Modulation      \\
Type        &   wavelength (nm) &   efficiency (\%) \\
\hline
B0         &   459.1            &   87.7            \\
A0         &   462.2            &   88.6            \\
F0         &   466.2            &   89.6            \\
G0         &   470.7            &   90.6            \\
K0         &   474.4            &   91.6            \\
M0         &   477.5            &   92.0            \\
M5         &   477.3            &   91.7            \\
\hline
\end{tabular}
\label{tab_eff}
\end{table}

Follow up observations were made for one object using SDSS $r^{\rm \prime}$ and 425 nm short pass (425SP) filters on 24/5/2015. The same procedures were used for calibrating telescope polarisation, angular calibration and efficiency as for the observations made using the SDSS $g^{\rm \prime}$ band. The $r^{\rm \prime}$ filter is 150 nm in width; it is centred on 625 nm but HIPPI's PMTs decrease in efficiency steadily from $\sim$525 nm, giving a shorter effective wavelength (this is depicted graphically in \citet{bailey15}). The response of the 425SP filter at the blue end is truncated by the detector response at $\sim$350 nm.

\subsection{Accounting for patchy cloud}
\label{subsec_patchy_cloud}

Some of the observations were made in cloudy conditions. For three objects (HIP 88635, HIP 107556 and to a lesser degree HIP 90496) the precision is considerably worse than the median (see Table \ref{tab_results}) owing to particularly poor weather on September 1st, 2014. Thick, rapidly moving patchy (mostly nimbus) cloud was constant for much of the night, and seeing was worse than 4$\arcsec$ at one point. Similarly patchy cloud also affected observations on September 2nd, but not to the same degree.

To account for this we removed the most cloud affected parts of the observations. The HIPPI observing technique allows for the efficient flagging and elimination of bad data. The data are taken in lots of 20 one second integrations per rotation. The Stokes parameters $Q$ or $U$ and $I$ are determined for each integration using a Mueller matrix method (see \citet{bailey15} for more details). The errors are then determined based on the number statistics of all the integrations in a rotation. Integrations with a Stokes $I$ determination of less than 5\% of the maximum observed for the corresponding $Q$ or $U$ Stokes parameter were discarded as cloud affected. The precision of the measurements is improved by removing the cloud affected integrations. However, each observation was initiated in as clear as possible conditions to ensure an appropriate maximum $I$.

The 5\% threshold was arrived at from very many repeated observations of HIP 2081 on September 1st, producing hundreds of integrations, made during the most varied observing conditions. By examining a moving average of one Stokes parameter with $I$, it was found that the accuracy of polarisation determinations was not affected, only the precision, at least down to a threshold of 3\%. 

\section{Results}

The resulting polarisation measurements for the 50 stars are given in Table \ref{tab_results}. This table lists the normalized Stokes parameters $q=Q/I$ and $u=U/I$, on the equatorial system, and the degree of polarisation and position angle obtained by combining the $q$ and $u$ measurements as $p=\sqrt{q^2+u^2}$. The polarisations and Stokes parameters are in units of parts per million (ppm, equal to 10$^{-6}$) in fractional polarisation.

The errors quoted are derived from the internal statistics of the individual data points included in each measurement as described in Section \ref{subsec_patchy_cloud} and by \cite{bailey15}. 

\begin{table*}
\caption{HIPPI linear polarisation measurements.}
\centering
\begin{tabular}{lllrrrr}
\hline
Star & Date(s)$^{a}$ & Cal$^{b}$ &
       \multicolumn{1}{c}{\hspace{3 mm}q (ppm)} & 
       \multicolumn{1}{c}{\hspace{3 mm}u (ppm)} & 
       \multicolumn{1}{c}{\hspace{2 mm}p (ppm)} &
       \multicolumn{1}{c}{\hspace{1.5 mm}$\theta$ ($\degr$)}\\  \hline
HIP 2021	&	28-31/8/14	& D &	$-$8.6 $\pm$ \hspace{1.5 mm}2.5	&	$-$1.6 $\pm$ \hspace{1.5 mm}2.5		&	8.8 $\pm$ \hspace{1.5 mm}2.5	&	95.1 $\pm$ 16.3	\\
HIP 2081	&	1/9/14	& S &	$-$5.4 $\pm$ \hspace{1.5 mm}4.6	&	$-$82.0 $\pm$ \hspace{1.5 mm}3.9		&	82.1 $\pm$ \hspace{1.5 mm}4.3	&	133.1 $\pm$ \hspace{1.5 mm}3.2	\\
HIP 3419	&	2/9/14	& S &	$-$23.1 $\pm$ \hspace{1.5 mm}8.0	&	$-$4.1 $\pm$ \hspace{1.5 mm}7.8		&	23.5 $\pm$ \hspace{1.5 mm}7.9	&	95.1 $\pm$ 19.1	\\
HIP 7588	& 2/9/14, 24/5/15 & D, S &	 969.6 $\pm$ \hspace{1.5 mm}4.0		&	 1920.9 $\pm$ \hspace{1.5 mm}3.6	&	2151.8 $\pm$ \hspace{1.5 mm}3.8	&	31.6 $\pm$ \hspace{1.5 mm}0.1	\\
HIP 9236	&	2/9/14	& D &	 31.7 $\pm$ \hspace{1.5 mm}8.2	&	$-$28.1 $\pm$ \hspace{1.5 mm}8.7		&	42.4 $\pm$ \hspace{1.5 mm}8.4	&	159.2 $\pm$ 11.5	\\
HIP 13847	&	2/9/14	& D &	 33.6 $\pm$ \hspace{1.5 mm}8.8	&	 65.9 $\pm$ \hspace{1.5 mm}8.0		&	74.0 $\pm$ \hspace{1.5 mm}8.4	&	31.5 $\pm$ \hspace{1.5 mm}6.7	\\
HIP 18543	&	31/8/14, 2/9/14 & D &	 42.0 $\pm$ \hspace{1.5 mm}6.7		&	 2.9 $\pm$ \hspace{1.5 mm}6.0		&	42.1 $\pm$ \hspace{1.5 mm}6.3	&	1.9 $\pm$ \hspace{1.5 mm}8.2	\\
HIP 26634	&	31/8/14	& S &	$-$668.4 $\pm$ \hspace{1.5 mm}6.5	&	$-$273.1 $\pm$ \hspace{1.5 mm}6.3	&	722.0 $\pm$ \hspace{1.5 mm}6.4	&	101.1 $\pm$ \hspace{1.5 mm}0.5	\\
HIP 30438	& 28-30/8/14, 2/9/14 & D &	$-$68.9 $\pm$ \hspace{1.5 mm}1.7	&	$-$89.2 $\pm$ \hspace{1.5 mm}1.6	&	112.8 $\pm$ \hspace{1.5 mm}1.7	&	116.2 $\pm$ \hspace{1.5 mm}0.9	\\
HIP 32349$^{c}$	& 31/8/14, 2/9/14, 23-24/5/15 &	D & $-$3.7 $\pm$ \hspace{1.5 mm}1.7	&	$-$4.0 $\pm$ \hspace{1.5 mm}1.7		&	5.5 $\pm$ \hspace{1.5 mm}1.7	&	113.8 $\pm$ 17.8	\\
HIP 39757	&	25/5/15	& S &	$-$13.4 $\pm$ \hspace{1.5 mm}5.8		&	$-$12.5 $\pm$ \hspace{1.5 mm}5.8		&	18.3 $\pm$ \hspace{1.5 mm}5.8	&	111.5 $\pm$ 18.2	\\
HIP 42913	&	25/5/15	& S &	$-$10.9 $\pm$ \hspace{1.5 mm}8.0		&	$-$43.1 $\pm$ \hspace{1.5 mm}8.0		&	44.5 $\pm$ \hspace{1.5 mm}8.0	&	127.9 $\pm$ 10.4	\\
HIP 45238	&	23/5/15	& S &	$-$20.5 $\pm$ \hspace{1.5 mm}3.3		&	 12.4 $\pm$ \hspace{1.5 mm}3.4		&	23.9 $\pm$ \hspace{1.5 mm}3.4	&	74.4 $\pm$ \hspace{1.5 mm}8.1	\\
HIP 46390	&	24/5/15	& S &	 8.4 $\pm$ \hspace{1.5 mm}4.7		&	$-$2.8 $\pm$ \hspace{1.5 mm}4.7		&	8.8 $\pm$ \hspace{1.5 mm}4.7	&	170.8 $\pm$ 30.3	\\
HIP 52727	&	12/5/14	& S &	$-$12.7 $\pm$ \hspace{1.5 mm}5.3		&	$-$30.4 $\pm$ \hspace{1.5 mm}4.6		&	33.0 $\pm$ \hspace{1.5 mm}4.9	&	123.7 $\pm$ \hspace{1.5 mm}9.0	\\
HIP 59803	&	12/5/14	& S &	 9.7 $\pm$ \hspace{1.5 mm}4.3		&	$-$68.2 $\pm$ \hspace{1.5 mm}3.4		&	68.9 $\pm$ \hspace{1.5 mm}3.9	&	139.1 $\pm$ \hspace{1.5 mm}3.5	\\
HIP 60718	&	12/5/14, 26/6/15	& S &	$-$248.1 $\pm$ \hspace{1.5 mm}2.3	&	$-$258.7 $\pm$ \hspace{1.5 mm}1.9	&	358.4 $\pm$ \hspace{1.5 mm}2.1	&	113.1 $\pm$ \hspace{1.5 mm}0.3	\\
HIP 60965	&	24/5/14	& S &	$-$14.3 $\pm$ \hspace{1.5 mm}4.8		&	 12.1 $\pm$ \hspace{1.5 mm}4.8		&	18.7 $\pm$ \hspace{1.5 mm}4.8	&	70.0 $\pm$ 14.7	\\
HIP 61084	&	12/5/14	& S &	$-$19.8 $\pm$ \hspace{1.5 mm}4.2		&	$-$39.8 $\pm$ \hspace{1.5 mm}3.5		&	44.5 $\pm$ \hspace{1.5 mm}3.9	&	121.8 $\pm$ \hspace{1.5 mm}5.3	\\
HIP 61359	&	12/5/14	& S &	$-$29.1 $\pm$ \hspace{1.5 mm}4.8		&	 15.3 $\pm$ \hspace{1.5 mm}4.1		&	32.9 $\pm$ \hspace{1.5 mm}4.4	&	76.1 $\pm$ \hspace{1.5 mm}7.4	\\
HIP 61585	&	12/5/14	& S &	 145.7 $\pm$ \hspace{1.5 mm}4.7		&	 3.3 $\pm$ \hspace{1.5 mm}3.4		&	145.7 $\pm$ \hspace{1.5 mm}4.0	&	0.7 $\pm$ \hspace{1.5 mm}1.3	\\
HIP 61932	&	24/5/15	& S &	$-$37.7 $\pm$ \hspace{1.5 mm}5.3		&	 48.8 $\pm$ \hspace{1.5 mm}5.1		&	61.6 $\pm$ \hspace{1.5 mm}5.2	&	63.8 $\pm$ \hspace{1.5 mm}4.9	\\
HIP 61941	&	24/5/15	& S &	$-$5.3 $\pm$ \hspace{1.5 mm}5.0		&	$-$5.4 $\pm$ \hspace{1.5 mm}5.1		&	7.6 $\pm$ \hspace{1.5 mm}5.0	&	112.5 $\pm$ 38.2	\\
HIP 64962	&	12/5/14	& S &	 2.2 $\pm$ \hspace{1.5 mm}5.1		&	 5.6 $\pm$ \hspace{1.5 mm}4.5		&	6.1 $\pm$ \hspace{1.5 mm}4.8	&	34.3 $\pm$ 47.9	\\
HIP 65109	&	2/9/14	& S &	$-$25.7 $\pm$ \hspace{1.5 mm}4.9		&	 22.2 $\pm$ \hspace{1.5 mm}4.7		&	34.0 $\pm$ \hspace{1.5 mm}4.8	&	69.6 $\pm$ \hspace{1.5 mm}8.0	\\
HIP 65474	&	24/5/15, 2$\times$29/6/15	& S &	$-$200.2 $\pm$ \hspace{1.5 mm}2.0	&	 42.7 $\pm$ \hspace{1.5 mm}2.0		&	204.7 $\pm$ \hspace{1.5 mm}2.0	&	84.0 $\pm$ \hspace{1.5 mm}0.6\\
HIP 68933	&	24/5/15	& S &	$-$16.9 $\pm$ \hspace{1.5 mm}4.3		&	 39.1 $\pm$ \hspace{1.5 mm}4.3		&	42.6 $\pm$ \hspace{1.5 mm}4.3	&	56.7 $\pm$ \hspace{1.5 mm}5.8	\\
HIP 71352	&	12/5/14	& S &	 5987.4 $\pm$ \hspace{1.5 mm}3.6		&	$-$1757.4 $\pm$ \hspace{1.5 mm}3.1	&	6240.0 $\pm$ \hspace{1.5 mm}3.4	&	171.8 $\pm$ \hspace{1.5 mm}0.0	\\
HIP 71683	&	12/5/14	& S &	 30.3 $\pm$ \hspace{1.5 mm}3.2		&	$-$30.0 $\pm$ \hspace{1.5 mm}2.1		&	42.6 $\pm$ \hspace{1.5 mm}2.7	&	157.6 $\pm$ \hspace{1.5 mm}3.7	\\
HIP 72622	&	12/5/14	& S &	 7.1 $\pm$ \hspace{1.5 mm}4.2		&	$-$26.6 $\pm$ \hspace{1.5 mm}3.6		&	27.5 $\pm$ \hspace{1.5 mm}3.9	&	142.5 $\pm$ \hspace{1.5 mm}8.6	\\
HIP 74785	&	12/5/14	& S &	 4.9 $\pm$ \hspace{1.5 mm}4.1		&	 150.5 $\pm$ \hspace{1.5 mm}3.4		&	150.6 $\pm$ \hspace{1.5 mm}3.7	&	44.1 $\pm$ \hspace{1.5 mm}1.6	\\
HIP 74946	&	12/5/14	& S &	 3.3 $\pm$ \hspace{1.5 mm}4.8		&	$-$29.4 $\pm$ \hspace{1.5 mm}4.1		&	29.5 $\pm$ \hspace{1.5 mm}4.4	&	138.2 $\pm$ \hspace{1.5 mm}9.2	\\
HIP 77952	&	9/2/14, 22/5/15 & S &	 5.3 $\pm$ \hspace{1.5 mm}4.4		&	 3.7 $\pm$ \hspace{1.5 mm}4.2		&	6.4 $\pm$ \hspace{1.5 mm}4.3	&	17.5 $\pm$ 37.9	\\
HIP 79593	& 12/5/14, 24/5/15 & S &	$-$50.1 $\pm$ \hspace{1.5 mm}4.7		&	$-$480.1 $\pm$ \hspace{1.5 mm}4.2	&	482.7 $\pm$ \hspace{1.5 mm}4.5	&	132.0 $\pm$ \hspace{1.5 mm}0.6	\\
HIP 82396	&	1/9/14	& S &	 10.5 $\pm$ \hspace{1.5 mm}6.6		&	 27.0 $\pm$ \hspace{1.5 mm}9.2		&	28.9 $\pm$ \hspace{1.5 mm}7.9	&	34.4 $\pm$ 13.8	\\
HIP 84012	&	12/5/14	& S &	 25.2 $\pm$ \hspace{1.5 mm}4.4		&	$-$50.7 $\pm$ \hspace{1.5 mm}3.8		&	56.6 $\pm$ \hspace{1.5 mm}4.1	&	148.2 $\pm$ \hspace{1.5 mm}4.4	\\
HIP 85792	&	12/5/14	& S &	 5113.8 $\pm$ \hspace{1.5 mm}4.0		&	$-$1469.1 $\pm$ \hspace{1.5 mm}3.5	&	5320.6 $\pm$ \hspace{1.5 mm}3.8	&	172.0 $\pm$ \hspace{1.5 mm}0.0	\\
HIP 86228	&	12/5/14	& S &	$-$149.3 $\pm$ \hspace{1.5 mm}3.6	&	$-$21.3 $\pm$ \hspace{1.5 mm}3.0		&	150.8 $\pm$ \hspace{1.5 mm}3.3	&	94.1 $\pm$ \hspace{1.5 mm}1.2	\\
HIP 88635	&	1/9/14	& S &	 37.6 $\pm$ 20.3		&	$-$8.3 $\pm$ 14.7		&	38.5 $\pm$ 17.5	&	173.8 $\pm$ 22.4	\\
HIP 89931	&	12/5/14	& S &	$-$313.1 $\pm$ \hspace{1.5 mm}5.1	&	$-$502.5 $\pm$ \hspace{1.5 mm}4.5	&	592.1 $\pm$ \hspace{1.5 mm}4.8	&	119.0 $\pm$ \hspace{1.5 mm}0.5	\\
HIP 90185	&	1/9/14	& S &	 38.8 $\pm$ \hspace{1.5 mm}4.5		&	 158.2 $\pm$ \hspace{1.5 mm}4.3		&	162.9 $\pm$ \hspace{1.5 mm}4.4	&	38.1 $\pm$ \hspace{1.5 mm}1.6	\\
HIP 90496	&	1/9/14	& S &	 9.6 $\pm$ 10.0		&	$-$53.4 $\pm$ \hspace{1.5 mm}9.1		&	54.2 $\pm$ \hspace{1.5 mm}9.5	&	140.1 $\pm$ 10.5	\\
HIP 92855	&	1/9/14, 22/5/15 & S &	$-$30.7 $\pm$ \hspace{1.5 mm}3.5		&	$-$166.9 $\pm$ \hspace{1.5 mm}3.4	&	169.7 $\pm$ \hspace{1.5 mm}3.4	&	129.8 $\pm$ \hspace{1.5 mm}1.2	\\
HIP 93506	&	1/9/14	& S &	 0.9 $\pm$ \hspace{1.5 mm}4.6		&	$-$28.2 $\pm$ \hspace{1.5 mm}4.7		&	28.2 $\pm$ \hspace{1.5 mm}4.6	&	135.9 $\pm$ \hspace{1.5 mm}9.3	\\
HIP 100751	&	1/9/14	& S &	$-$5.7 $\pm$ \hspace{1.5 mm}4.8		&	$-$85.5 $\pm$ \hspace{1.5 mm}4.1		&	85.6 $\pm$ \hspace{1.5 mm}4.4	&	133.1 $\pm$ \hspace{1.5 mm}3.2	\\
HIP 107556	&	1/9/14	& S &	 2.4 $\pm$ 19.8		&	$-$29.5 $\pm$ 15.6	&	29.6 $\pm$ 17.7	&	137.3 $\pm$ 38.3	\\
HIP 109268	&	31/8/14	& D &	$-$91.4 $\pm$ \hspace{1.5 mm}3.4		&	 21.1 $\pm$ \hspace{1.5 mm}3.1		&	93.8 $\pm$ \hspace{1.5 mm}3.3	&	83.5 $\pm$ \hspace{1.5 mm}1.9	\\
HIP 110130	&	31/8/14, 26/6/15 & D, S &	$-$107.4 $\pm$ \hspace{1.5 mm}1.8	&	$-$74.7 $\pm$ \hspace{1.5 mm}4.2		&	131.1 $\pm$ \hspace{1.5 mm}4.2	&	107.4 $\pm$ \hspace{1.5 mm}1.8	\\
HIP 112122	&	1/9/14	& S &	 330.7 $\pm$ \hspace{1.5 mm}5.8		&	 560.2 $\pm$ \hspace{1.5 mm}8.1		&	650.5 $\pm$ \hspace{1.5 mm}7.0	&	29.7 $\pm$ \hspace{1.5 mm}0.6	\\
HIP 113368	&	28/8/14	& D &	$-$17.8 $\pm$ \hspace{1.5 mm}3.2		&	$-$16.6 $\pm$ \hspace{1.5 mm}3.0		&	24.3 $\pm$ \hspace{1.5 mm}3.1	&	111.5 $\pm$ \hspace{1.5 mm}7.4	\\
\hline
\end{tabular}
\begin{flushleft}
a - Hyphenation indicates object was observed once per day inclusive. \\
b - Calibration type: full sky subtraction (S), or dark subtraction only (D). \\
c - Sirius was used as a low polarisation standard.
\end{flushleft}
\label{tab_results}
\end{table*}

\subsection{Uncertain results for binaries with aperture scale separations}
\label{sec_binary_ap_sep}

The process of centring a star in HIPPI's 6.7$\arcsec$ aperture is carried out by manual scanning to maximise the total signal received by the instrument. This process is made difficult when stars in a binary system have a separation similar to the radius of the aperture, particularly if they have a similar apparent magnitude, or if the seeing is poor. This can result in the system being off-centre in the aperture, and a partial contribution from the secondary. In such instances a small instrumental polarisation is induced that would be difficult to calibrate for.

\subsubsection{$\alpha$ Cen}
\label{sec_alpha_cen}

This particular difficulty was apparent when attempting to acquire $\alpha$ Cen A (HIP 71683). The separation of $\alpha$ Cen A (G2V) and B (K1V) at the time of our observations was around 4$\arcsec$ (the separation is depicted graphically by \citet{burnham78a}, and the updated parameters given by \citet{pourbaix02} are little different). We certainly have a significant contribution from component B also, and so we have reported the polarisation for the $\alpha$ Cen system as a whole. In the discussion that follows it will become clear that the degree of polarisation of $\alpha$ Cen is anomalously high when one considers its spectral type and proximity to the Sun, and we ascribe this to aperture scale separation of $\alpha$ Cen B inducing instrumental polarisation.

\subsubsection{$\alpha$ Cru}
\label{sec_alpha_cru}

Although not apparent at the time, it is possible that $\alpha$ Cru (HIP 60718) is affected in the same way. The components $\alpha^1$ (B0.5IV) and $\alpha^2$ (B1V) are also separated by $\sim$4$\arcsec$ \citep{burnham78b,pourbaix04}. $\alpha$ Cru shows higher polarisation than any other (non-Be) B type star in our survey which suggests that its measurement is spurious. However, there are other factors that may result in a high degree of polarisation for $\alpha$ Cru. It is the star system in the survey with the earliest spectral type, and the $\alpha^1$ component is itself a binary with a separation of $\sim$1 AU, while $\alpha^2$ may also be a binary \citep{burnham78b,pourbaix04}. These factors are discussed with reference to other stars in Sections \ref{sec_b_stars} and \ref{sec_binary}. 

\section{Discussion}

The discussion begins with a comparison to previous results (\ref{sec_prev}) and a look at the statistics of the survey (\ref{sec_stats}). Thereafter it is divided up roughly by stellar spectral type. Polarigenic mechanisms for A-K stars begin with the spatial distribution of polarisation due to the interstellar medium (\ref{sec_spatial}), we then look at debris disk systems (\ref{sec_debris_disk}), Ap stars (\ref{sec_ap}) and eclipsing binaries (\ref{sec_eclipse}). B spectral types are examined next beginning with close binaries ({\ref{sec_binary}), then Be stars (\ref{sec_be_stars}) before concluding with a look at the remainder of B stars ({\ref{sec_b_stars}). Finally we consider the late giants in the survey (\ref{sec_giants}).

\subsection{Comparison with previous observations}
\label{sec_prev}

Of the stars highlighted in Section \ref{sec_sample} as having previous significant polarisation measurements, all have significant measurements in our survey also. Our determinations are all of the same order of magnitude as the previous ones. However, all are significantly different. This is reflective either of the improved precision of HIPPI over older instruments, or these three stars are variable in polarisation -- an indication that the polarisation is intrinsic. A less likely possibility is that the variation represents the movement of the stars with respect to the patchy interstellar medium, but the variability we see would correspond to the movement of many tens of parsecs worth of dust based on the polarisation with distance relation found by \citet{bailey10} for northern stars.

The best match for previous measurements is with that of \citet{tinbergen82} for Spica, where we are in agreement within the error for both previous measurements in $q$ and significantly different to less than 2$\sigma$ in $u$. The measurement tabulated by \citet{heiles00} is half the degree of polarisation and 63$\degr$ different. Spica will be discussed in detail in Section \ref{sec_spica}.

For the late Giant $\delta$ Sgr we are in agreement with \citet{tinbergen82} in $u$, but have a thrice greater $q$ measurement that is significant to more than 3$\sigma$. For $\beta$ Gru we have double the polarisation of \citet{heiles00} and a polarisation angle that is 126$\degr$ different, again this is significant to more than 3$\sigma$. This is strong evidence for variable intrinsic polarisation. These two stars will be discussed in Section \ref{sec_giants}.

\subsubsection{Sirius}

Like \citet{bailey15} we have used Sirius as a low polarisation standard, and in fact combined the measurements reported there with additional observations. We now make its polarisation as 5.5$\pm$1.7 ppm. From the discussion below in Section \ref{sec_spatial} it will be seen that this is entirely consistent with interstellar polarisation at southern declinations. At 2.6 pc distant and a V magnitude of -1.46 Sirius is a particularly good unpolarised standard for instruments that can tolerate its brilliance. It should be pointed out though that the white dwarf companion, Sirius B, is currently separated from the primary by 10.2$\arcsec$ \citep{burnham78a} and well outside the 6.7$\arcsec$ HIPPI aperture. Care will need to be taken with larger instrument apertures or closer to periastron (which next occurs in 2044) when the separation is 3$\arcsec$ \citep{burnham78a}.

\subsection{Survey statistics}
\label{sec_stats}

\begin{table}
\caption{Polarisation for spectral classes.}
\centering
\begin{tabular}{l r r r r r r}
\hline
Spectral & \multicolumn{3}{c}{HIPPI$^a$} & \multicolumn{3}{c}{PlanetPol$^b$} \\
Class    & N      & Mean        & Median       & N     & Mean       & Median    \\
         &        & d (pc)   & p (ppm)    &       & d (pc)  & p (ppm) \\  
\hline
B$^{c}$  &    9   &  65.0       &  145.7       &  5    &   49.6     &  36.8      \\
A        &   14   &  31.7       &   29.6       & 18    &   28.1     &   8.4      \\
F/G      &   10   &  45.5       &   32.9       &  9    &   23.5     &   9.4      \\
K        &    9   &  40.9       &   54.2       & 12    &   38.7     &   9.8      \\
M        &    4   &  49.1       &  263.6       &  4    &   78.6     & 109.2      \\
\hline
\end{tabular}
\begin{flushleft}
a - This work. \\
b - \cite{bailey10}. \\
c - Be stars not included for HIPPI; if included: 13, 67.3, 166.6. \\
\end{flushleft}
\label{tab_spc}
\end{table}

In Table \ref{tab_spc} the polarisation properties as a function of spectral type for both this work and the survey conducted by \citet{bailey10} with PlanetPol are listed for comparison. The PlanetPol instrument operated with a very broadband red filter covering the wavelength range from 590 nm to the detector cutoff at about 1000 nm \citep{hough06} -- redder than our measurements made with HIPPI. The empirical wavelength dependence of interstellar polarisation is given by \textit{Serkowski's Law} \citep{wilking82}: \begin{equation} \label{eq_serkowski} p(\lambda/\lambda_{max})=\exp((0.1-1.86\lambda_{max})\ln^{2}(\lambda/\lambda_{max})),\end{equation} where $\lambda$ is the wavelength examined and $\lambda_{max}$ the wavelength of maximum polarisation. A typical value for $\lambda_{max}$ is 550 nm \citep{serkowski75}, and for the effective wavelength of a G0 star observed by HIPPI and PlanetPol this corresponds to a factor 1.08 times greater polarisation for HIPPI. However, extremes in $\lambda_{max}$ can range from 340 to 1000 nm. The difficulty in measuring polarisation to the ppm level has so far prevented $\lambda_{max}$ from being determined within 100 pc.

Before analysing Table \ref{tab_spc} it is pertinent to briefly discuss the statistics of the degree of polarisation in such a survey as this. The degree of polarisation of a system, $p$, is the vector sum of intrinsic, $p_*$, and interstellar, $p_i$, components; from the Law of Cosines: \begin{equation} \label{eq_p_stat} p=\sqrt{p_i^2+p_*^2-2p_ip_*\cos(\pi-2\alpha)},\end{equation} where $\alpha$ is the angle of interstellar polarisation relative to the direction of the intrinsic component. As we do not know the orientation of either component of polarisation, $\alpha$ will have a random value between 0 and $\upi$. Figure \ref{fig_p_stat} (a) shows how this effects polarisation measured as a function of the ratio of intrinsic to interstellar polarisation. If the intrinsic component of polarisation is more than twice the interstellar component, then the polarisation measured will always be greater than $p_i$. However, even for relatively small values of $p_*$, on the average we expect $p$ to be greater than $p_i$, even if a fraction of polarisations measured are less than the degree of interstellar polarisation -- as shown in Figure \ref{fig_p_stat} (b).

\begin{figure}
\centering
\includegraphics[width=0.5\textwidth,natwidth=100,natheight=100,trim=0.5cm 1.5cm 0.5cm 0cm]{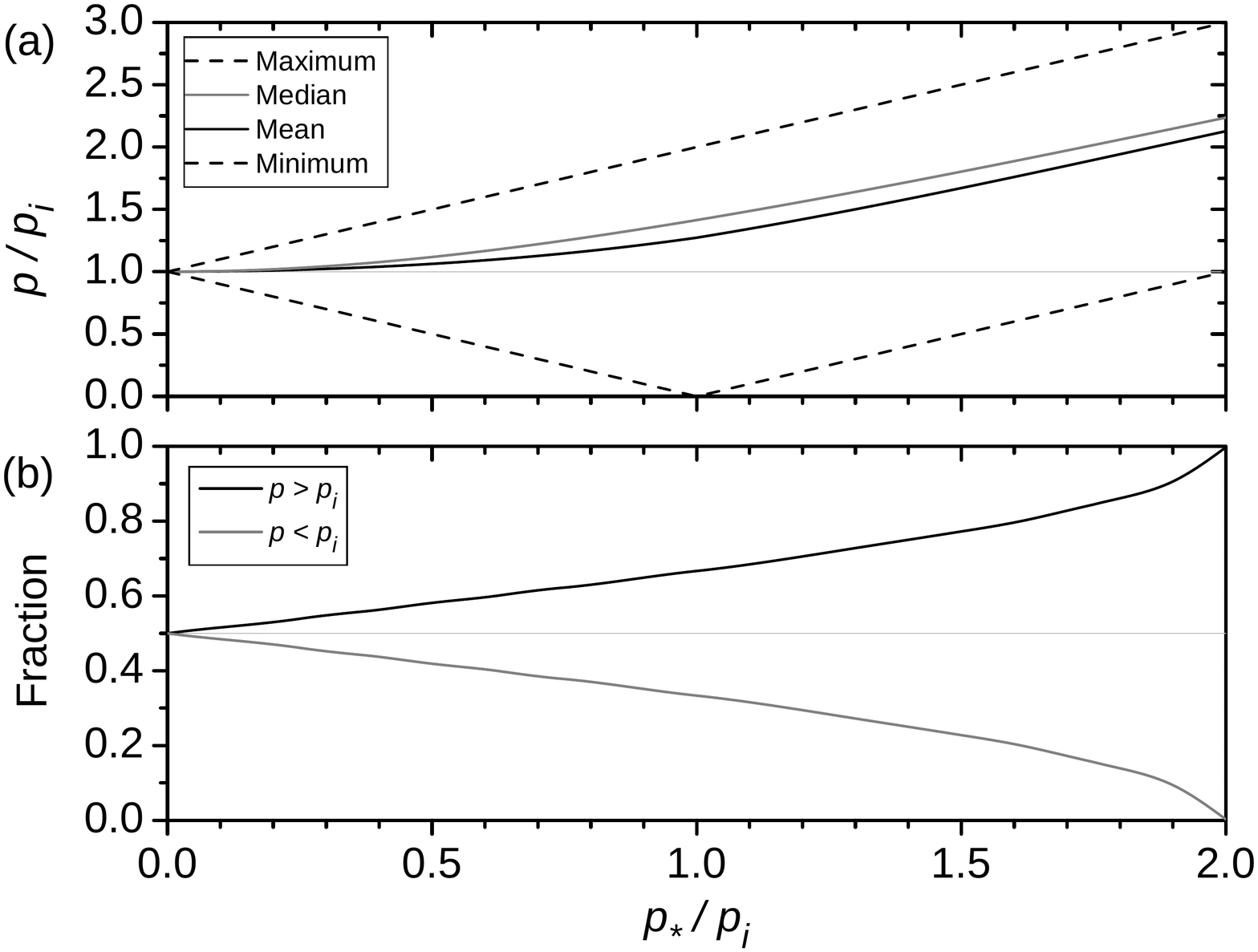}
\caption{(a) How the polarisation measured, $p$, changes as the ratio of intrinsic to interstellar components, $p_*/p_i$. The range of values between the plotted maximum and minimum (dashed) are possible, but for a sufficiently large sample the mean (black) and median (grey) $p$ will always greater than $p_i$ for a non-zero $p_*$. (b) As $p_*/p_i$ increases the probability of a random measurement of $p$ for a star being greater than $p_i$ increases.}
\label{fig_p_stat}
\end{figure}

From Table \ref{tab_spc} the median polarisation is similar across the spectral classes A, F, G and K for both instruments. \citet{tinbergen82} suspected the presence of variable intrinsic polarisation at the 10$^{-4}$ level in stars with spectral type F0 and later. This supposition was not supported by the more sensitive observations of \citet{bailey10}. Higher polarisations were indicated for spectral classes B and M, but they had very few stars of those types, and at larger average distances, and so could not ascribe that to any intrinsic polarisation of those stars. 

Adding our results to those obtained with PlanetPol makes it clear that B and M spectral classes do have higher polarisations; this is true even if we discount the Be stars. As is to be expected from our selected V magnitude limit of 3.0 the B stars are on average at larger distances than those of other spectral classes -- more than twice the mean distance than for A stars -- yet the median polarisation is more than 4 times that of the A stars. We have fewer M stars, together with the PlanetPol only eight, but three of them have a polarisation of $\sim$500 ppm or greater. Additionally, the furthest K giant (K3III), $\delta$ Sgr, also has a polarisation in excess of 500 ppm. As will be discussed later, the interstellar contribution is not likely to be much more than 150 ppm even at 100 pc, and so a measurement of 500 ppm, in addition to the variability observed with respect to the observations of \citet{tinbergen82}, strongly argues for an intrinsic cause.

\begin{figure*}
\centering
\includegraphics[width=1.0\textwidth,natwidth=600,natheight=400,trim=0.5cm 3cm 0.5cm 2cm]{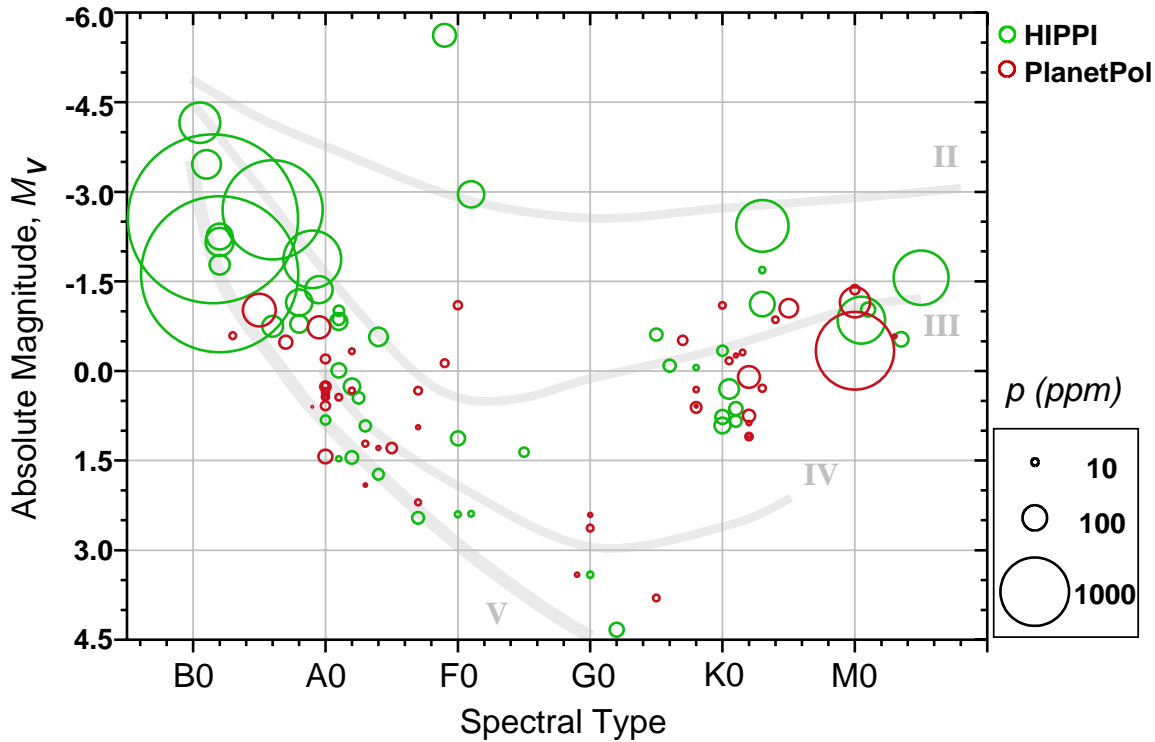}
\caption{The results of both the HIPPI (green) and PlanetPol (red) surveys plotted on a H-R Diagram. The degree of polarisation detected is represented by the area of the bubble. The main sequence (V), sub-giant (IV), giant (III) and bright giant (II) branches are indicated. Two clear trends are observed: (1) increased polarisation of B stars, and (2) increased polarisation of late giant stars.}
\label{fig_HR}
\end{figure*}

Plotting the polarisation for all stars observed by us and \citet{bailey10} on a H-R Diagram (Figure \ref{fig_HR}) best illustrates the trends with spectral type. Even if one ignores the highest polarisations of the four Be stars, it can be seen that there is a stark contrast around spectral type A0, where earlier types have a greater degree of polarisation. The sharpness of the division between A and B stars suggests an intrinsic mechanism or mechanisms particular to this stellar type; this will be discussed in Section \ref{sec_b_stars}. 

On the giant branch (luminosity class III) we have very sparse data earlier than G5. For later types, and greater luminosities, there is a trend toward higher polarisations. In Section \ref{sec_giants} we identify a number of K and M giants that we can be reasonably certain are intrinsically polarised and investigate the most likely mechanisms.

\subsection{Polarisation spatial distribution}
\label{sec_spatial}

To investigate the spatial distribution of interstellar polarisation we have plotted PlanetPol and HIPPI survey stars' polarisation degree and angle in equatorial co-ordinates (Figure \ref{fig_Equ}) and projected on to Galactic co-ordinates (Figure \ref{fig_Gal}). B type stars, M type stars and the K giant $\sigma$ Sgr have been neglected to reduce contamination from intrinsic polarisation.

The PlanetPol survey \citep{bailey10} found low interstellar polarisation toward the Galactic north pole, with higher polarisations below 40$\degr$ Galactic latitude toward the Galactic centre\footnote{It should be pointed out that the spatial scale of measurements made by PlanetPol and HIPPI is many orders of magnitude less than the Galactic scale, and that in this case a maximum seen toward the Galactic centre is coincidental. The Solar System is currently located between the Orion Spur and the Sagittarius Main Arm.}, and that in this range there was a tendency for alignment in the Galactic plane. Removing the B and M stars hasn't changed that picture (Figure \ref{fig_Gal}). The new results we present here align well with those of PlanetPol. The few data points we add at high Galactic latitudes are relatively low polarisation. Near the Galactic centre we add data at lower northern Galactic latitudes; the polarisation vectors for these data points between \textit{l} = $\sim$0-120$\degr$ show an alignment consistent with the PlanetPol data, i.e. along the Galactic plane. Further south a different picture emerges.

Our survey has added significantly more data points in the Galactic south. Even if one allows for the wavelength difference between HIPPI using the $g^{\rm \prime}$ filter and PlanetPol's 600-800 nm window, in general southern stars appear significantly more polarised than northern ones. That the degree of polarisation is greater in the Galactic south than north is consistent with the distribution of interstellar dust clouds seen around O, B, F and G stars by IRAS at 60 $\mu$m \citep{dring96}. This might be explained by reference to the Sun's vertical displacement from the Galactic plane, \textit{z$_\odot$}. Many studies have shown that the Sun is slightly to the north of the Galactic plane compared to its assumed position in the Galactic co-ordinate system \citep{joshi07}. Thus, by observing to the south, we are looking through more of the Galactic plane where we might expect interstellar grains to be more strongly aligned and impart greater polarisation.

By plotting the polarisation degree per distance in Figure \ref{fig_EquPdist} and with distance in Figure \ref{fig_Pwd} we can refine the picture further. We denote the polarisation with distance as $p/d$, it is useful for comparing the degree of polarisation of objects that might be separated by tens of pc. In Figures \ref{fig_EquPdist} and \ref{fig_Pwd}, the polarisation has been debiased as $\sqrt{p^2-\sigma^2}$. \citet{bailey10} noted at right ascensions greater than 17 h a $p/d$ relation of $\sim$2$\times$10$^{-6}$ pc$^{-1}$. With the removal of the B and M stars from the map, there remain only 3 stars supporting this trend for declinations greater than 30$\degr$. One of these is Vega, which we can discount from considerations of interstellar polarisation on the basis of its large debris disk. The remaining two stars are the K giants $\gamma$ Dra and $\kappa$ Lyr. From the discussion that follows in Section \ref{sec_giants} it is likely that these stars are also intrinsically polarised. Indeed, cross-reference between Figures \ref{fig_Equ} and \ref{fig_EquPdist} reveals the position angles of these three stars do not match those nearby. Given that they all lie near the Galactic equator, where there is otherwise a good measure of alignment, the case is strong for intrinsic polarisation.

The large degree of polarisation observed for $\alpha$ Cen is probably not real, and is discussed in detail in Section \ref{sec_binary_ap_sep}. From our survey, two further K giants, $\alpha$ Phe and $\alpha$ Tuc, can be identified as probably intrinsically polarised, as they have a $p/d$ greater than any other stars within 100 pc. What remains is a region of low $p/d$ centred on 14 h right ascension, +35$\degr$ declination -- this region was mostly covered by the PlanetPol survey -- indicating a relatively dust free volume. Within this region stars are polarised at $\sim$2$\times$10$^{-7}$ pc$^{-1}$. Further South and East $p/d$ is greater, and somewhat patchy. The stars with the greatest $p/d$ tend to lie within -15$\degr$ and -30$\degr$ declination or nearby; these stars are polarised at $\sim$2$\times$10$^{-6}$ pc$^{-1}$. Between the two regions are stars with intermediate polarisations. A linear fit to stars from our survey not suspected of being intrinsically polarised gives 1.14$\times$10$^{-6}$ pc$^{-1}$ with a coefficient of determination, $R^2=$ 0.67. The highly polarised southern stars for the most part lie within 30 pc. Though we have fewer stars further away, these are mostly less polarised. This suggests that the higher interstellar polarisation of southern stars is imparted predominantly by nearby dust lying, from Figure \ref{fig_Pwd}, between 10 and 30 pc distant. Coincidentally, the majority of estimates for \textit{z$_\odot$} place the Sun 15 to 30 pc above the Galactic plane \citep{joshi07}, though the implication is most likely a local dust cloud at this distance.

The polarisation of more distant stars in the interstellar medium has been shown to increase with distance as $\sim$2x10$^{-5}$ pc$^{-1}$ \citep{behr59}. That nearby stars show less polarisation was demonstrated by \citet{tinbergen82} and \citet{leroy93b}. Exactly how much less was determined by \citet{bailey10} for northern stars in the PlanetPol survey. \citet{bailey10} believed that the furthest stars in their survey showed high polarisations as a result of their proximity to the wall of the Local Bubble. As a result of the work presented here, we now believe those stars to be polarised by intrinsic processes. The two furthest stars shown in the figures are Canopus and $\theta$ Sco from our survey; they exhibit the greatest degree of polarisation. This may be due to these stars being beyond the wall of the Local Bubble. Repeat observations of Canopus don't show any variation \citep{bailey15}, which supports this. However, they are also the most luminous non-B stars in the survey, and to the best of our knowledge the polarisation of close bright giants at the ppm level has not previously been investigated (though it should be noted that early type supergiants do display intrinsic aperiodic variable polarisation arising as a result of asymmetric mass loss \citep{hayes84,hayes86}). The 3rd magnitude limit in V for this survey has resulted in fewer main sequence and sub-giant stars at distance than the PlanetPol survey with its limit of 5th magnitude. Without more data it isn't possible to say definitively whether we're probing the wall of the Local Bubble with our furthest stars.  

\begin{figure*}
\centering
\includegraphics[width=0.8\textwidth,natwidth=600,natheight=400,trim=0.5cm 0.5cm 0.5cm 0.5cm]{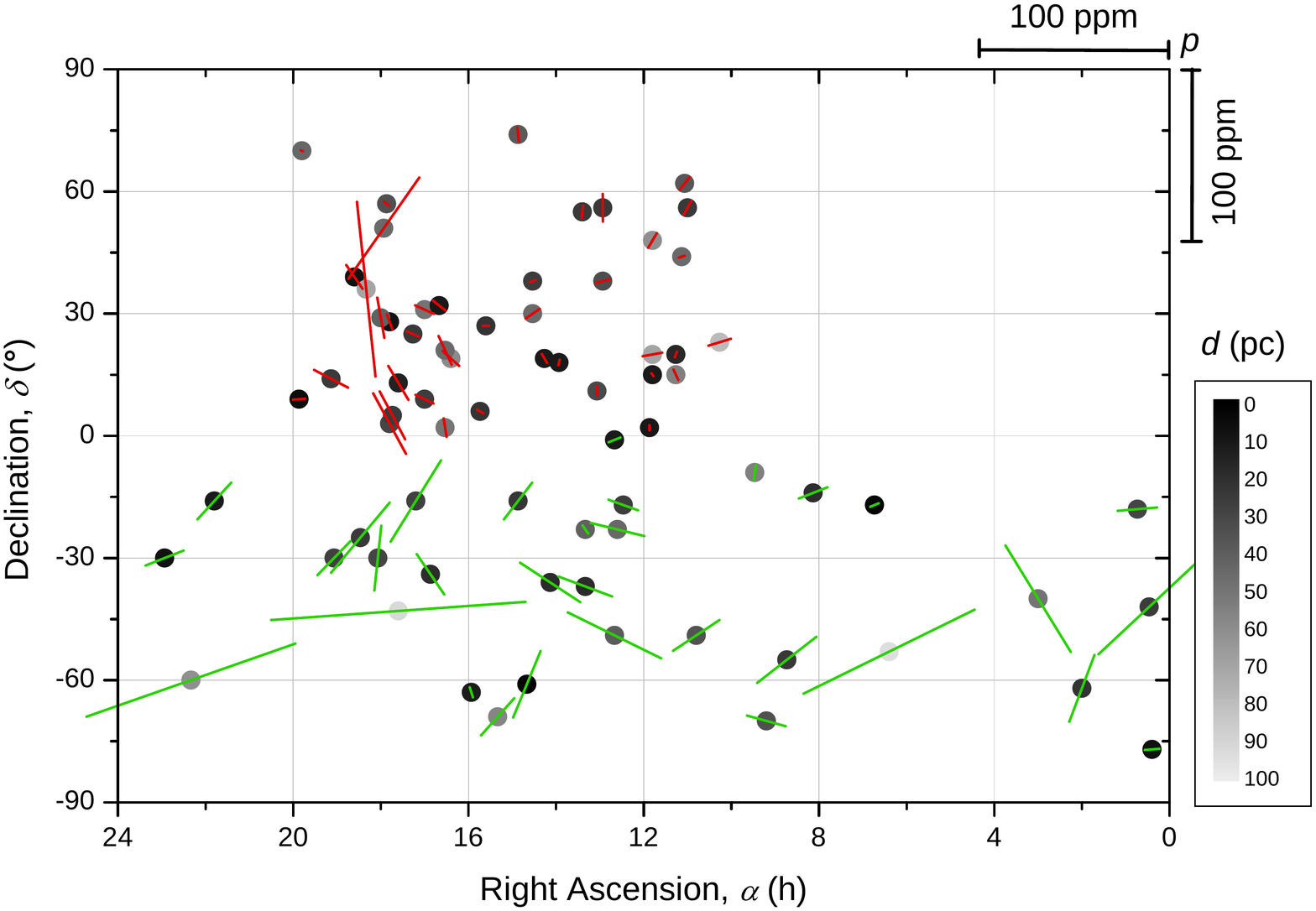}
\caption{A to K stars within 100 pc from the HIPPI (green) and PlanetPol (red) surveys plotted on an equatorial co-ordinate system. Stellar positions are marked by dots, with a greyscale indicating distance from the Sun in pc. Vectors denote the degree and angle of the polarisation.}
\label{fig_Equ}
\end{figure*}

\begin{figure*}
\centering
\includegraphics[width=0.8\textwidth,natwidth=600,natheight=400,trim=0.5cm 0.5cm 0.5cm 0.5cm]{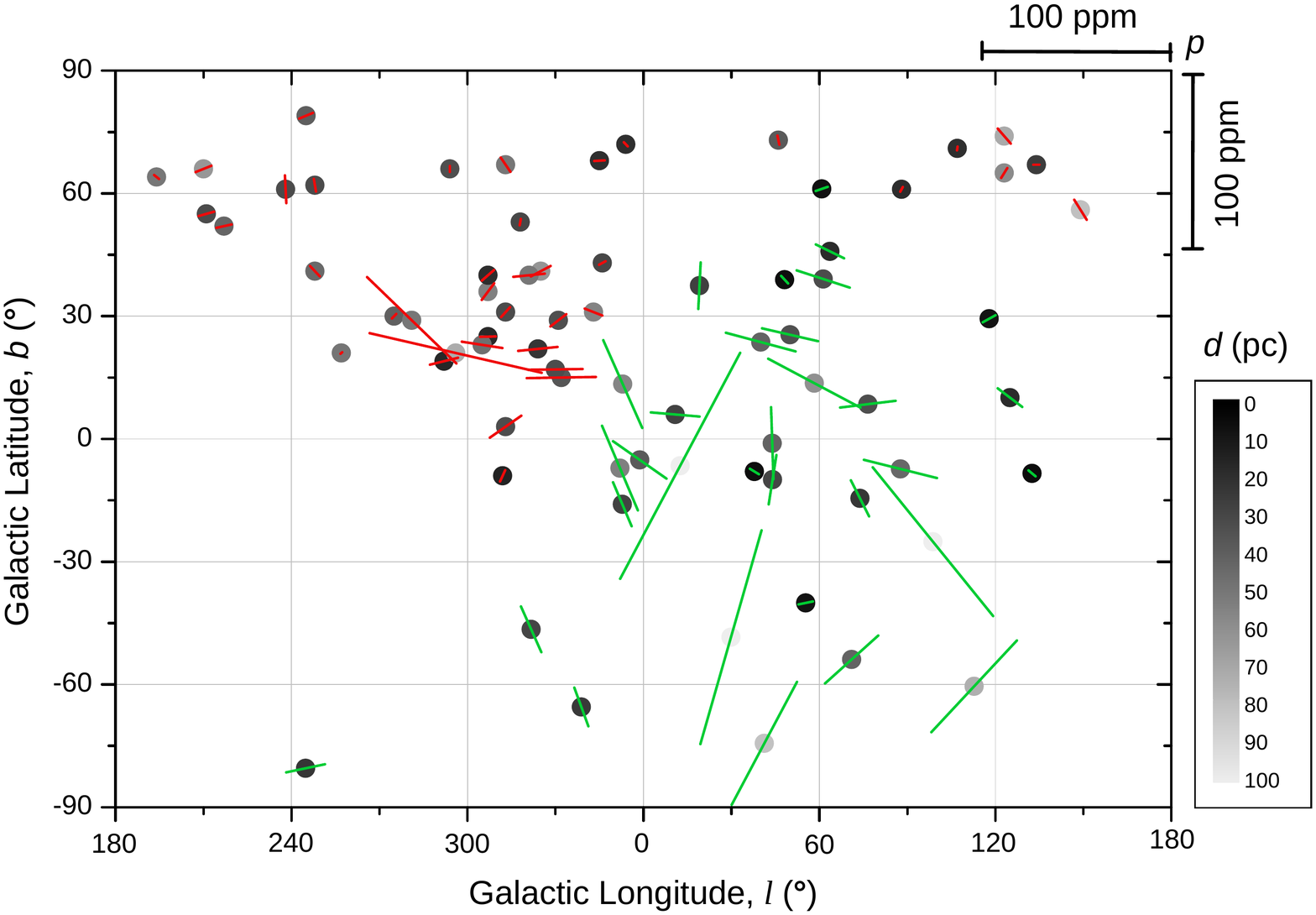}
\caption{A to K stars within 100 pc from the HIPPI (green) and PlanetPol (red) surveys plotted on Galactic co-ordinates. Stellar positions are marked by dots, with a greyscale indicating distance from the Sun in pc. Vectors denote the degree and angle of the polarisation.}
\label{fig_Gal}
\end{figure*}

\begin{figure*}
\centering
\includegraphics[width=0.8\textwidth,natwidth=600,natheight=400,trim=0.5cm 0.5cm 0.5cm 0.5cm]{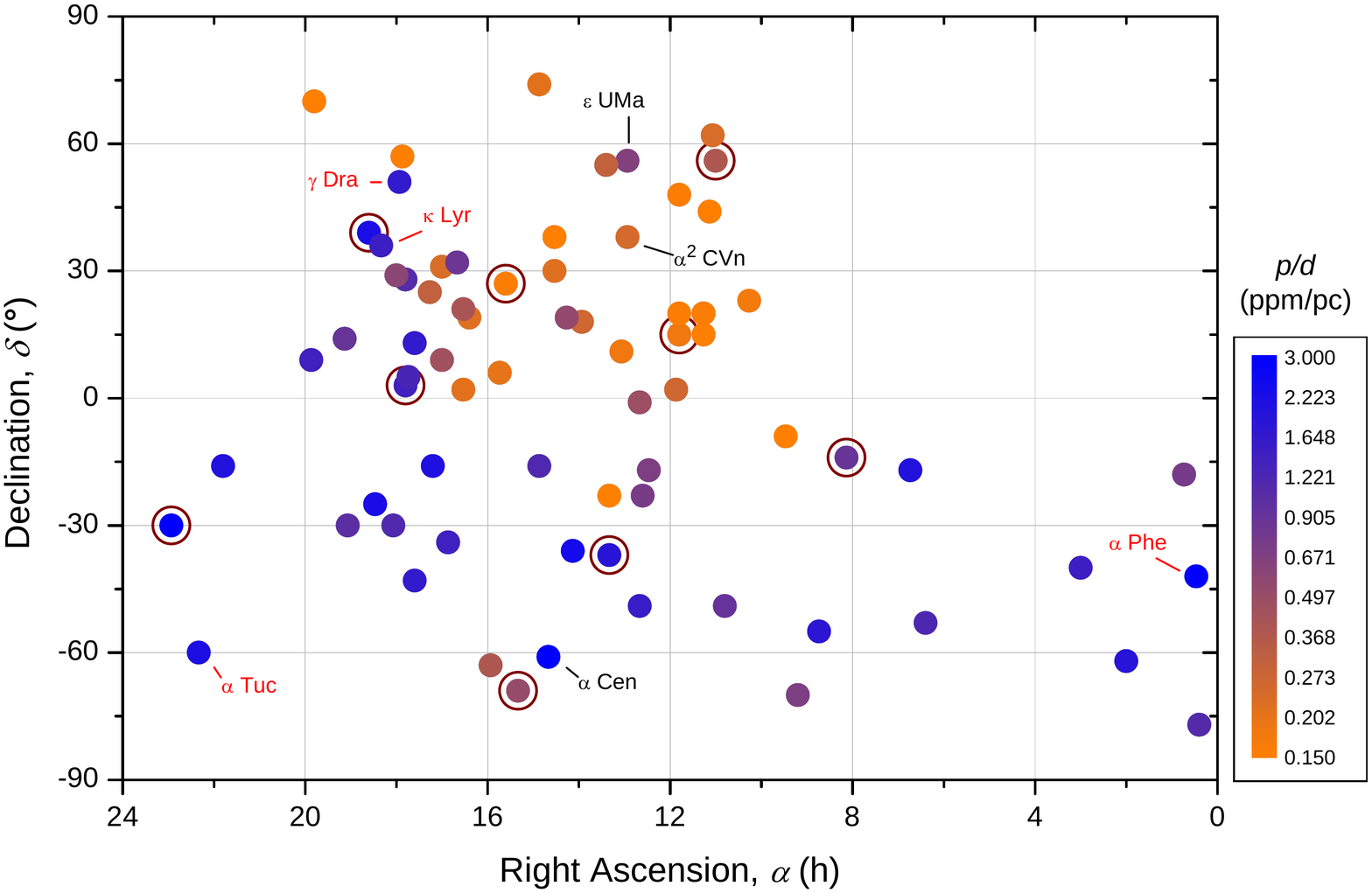}
\caption{A to K stars within 100 pc from the HIPPI (green) and PlanetPol (red) surveys plotted on an equatorial co-ordinate system. Stellar positions are marked by dots. Polarisation with distance, $p/d$, in ppm/pc is given by the colour scale to align with Figure \ref{fig_Pwd}. Brown rings denote debris disk stars. Stars labelled in red are K giants that may be intrinsically polarised. Stars labelled in black may not be representative of interstellar polarisation for other reasons (see the text).}
\label{fig_EquPdist}
\end{figure*}

\begin{figure*}
\centering
\includegraphics[width=0.8\textwidth,natwidth=600,natheight=400,trim=0.5cm 0.5cm 0.5cm 0.5cm]{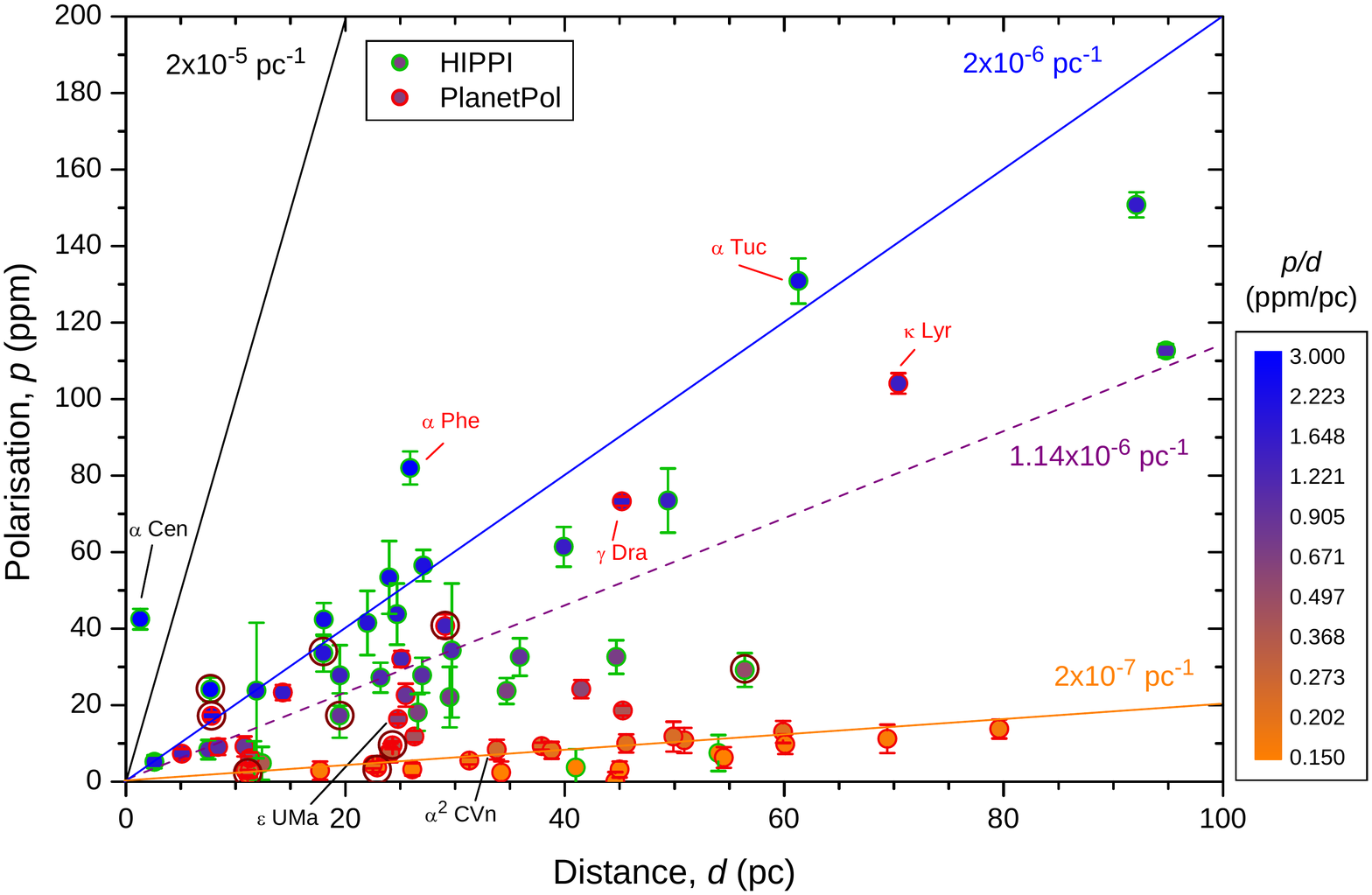}
\caption{Polarisation with distance, $p/d$, for A to K stars within 100 pc from the HIPPI (green) and PlanetPol (red) surveys. Stellar positions are marked by dots. Polarisation with distance in ppm/pc is given by the colour scale to align with Figure \ref{fig_EquPdist}. Brown rings denote debris disk stars. Stars labelled in red are K giants that may be intrinsically polarised. Stars labelled in black may not be representative of interstellar polarisation for other reasons (see the text). The dashed line represents a linear fit to the unannoted HIPPI stars.}
\label{fig_Pwd}
\end{figure*}

\subsection{Debris disks}
\label{sec_debris_disk}

Debris disks are circumstellar disks of dust around main-sequence stars. They are typically detected and characterised by excess infrared emission above that of the stellar photospheric continuum emission. The peak wavelength of dust emission from debris disks depends primarily on the distance from the star \citep{matthews14}. Broadly speaking, two typical temperature regimes are observed; warm dust ($T_d$ 150 to 220 K) analogous to the Asteroid belt, and cold dust ($T_d$ $\sim$ 50 to 80 K) analogous to the Edgeworth-Kuiper belt \citep{morales11}. Edgeworth-Kuiper belt analogues are more easily detected by this method than closer disks owing to the greater contribution of the stellar photosphere to the total emission at mid-infrared wavelengths. Recent far-infrared surveys with the PACS instrument \citep{poglitsch10} on the \textit{Herschel Space Observatory} \citep{pilbratt10}, where detection is already sensitivity limited, have found debris disks for up to 33\% of A stars \citep{thureau14}, and 20\% of FGK stars \citep{eiroa13}. The ratio of the luminosity of the disk to that of the star, $F_{\rm IR}/F_{\star}$, is indicative of the total starlight intercepted by the debris disk \citep{wyatt08}, but by itself is not a reliable predictor of scattered light brightness \citep{schneider14b}. The spectral energy distribution of a debris disk is also a function of the dust particle size distribution and composition as well as its architecture. Dust particles in circumstellar disks polarise light by scattering and absorption processes, and sensitive polarimetry is potentially useful for removing degeneracies \citep{garcia15,schneider14b}. Polarisation seen by aperture polarimetry -- where the aperture takes in the central star as well as the whole/a large portion of the disk -- has been reported at levels of $\sim$0.1 to 2\% (\citet{garcia15} and references therein).

As part of our survey we observed four main sequence objects thought to be debris disk host stars, namely: Fomalhaut (HIP 113368), $\iota$ Cen (HIP 65109), $\beta$ TrA (HIP 77952) and $\gamma$ TrA (HIP 74946). The PlanetPol survey also observed five debris disk systems: Vega (BS 7001), Merak (BS 4295), $\gamma$ Oph (BS 6629), $\beta$ Leo (BS 4534) and $\alpha$ CrB (BS 5793). Concurrently with the bright star survey we carried out another program where we observed debris disk systems. What can be learned from all of these observations has been substantially dealt with in another paper \citep{marshall15}. In brief, comparison of the polarisation with measurements of the disks' thermal emission demonstrate the capacity for polarisation measurements to constrain the geometry and orientation of unresolved debris disks. A typical ratio of polarisation to thermal emission of between 5 to 50\%, consistent with scattered light imaging measurements. The majority of disks in the sample have polarisation signals aligned approximately perpendicular to the disk major axis, indicative of scattering from the limb of small grains in those disks. No trend was found in the polarisation with either disk inclination, nor luminosity of the host star. The characterisation of the system $\beta$ TrA as debris disk system based on a far infrared excess derived from a single waveband \textit{IRAS} measurement is found to be unreliable.

What remains to be presented here is a comparison of polarisation in these debris disk systems compared with those of the other main sequence stars in the survey, this is shown in Table \ref{tab_dd}. If the contribution from the disk is greater than the interstellar polarisation then it is obvious that we would expect a higher polarisation on average from debris disk systems. Yet geometrical considerations related to the vector sum of intrinsic and interstellar components mean we should also expect a slightly higher polarisation from debris disk systems on average when the magnitude of contributions is similar or even smaller (as detailed in Figure \ref{fig_p_stat}). The situation is depicted graphically in Figure \ref{fig_vector}.

\begin{figure}
\includegraphics[width=0.5\textwidth,natwidth=100,natheight=100,trim=3.5cm 4cm 7cm 2cm]{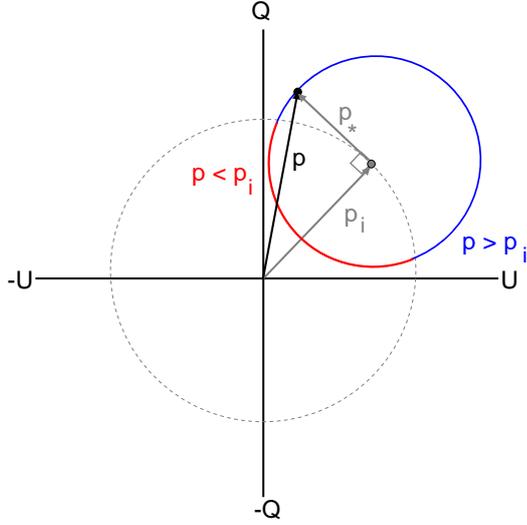}
\caption{The vector sum, $p$, of interstellar, $p_{i}$, and intrinsic, $p_{*}$, components of polarisation shown in the Q-U plane. The situation where the sum is less than the interstellar polarisation by itself is represented by the red arc; the situation where it is greater is represented by the blue arc. Given a random distribution more points will lie on the blue arc than the red and thus the average $p$ will be greater than $p_{i}$. The polarisation is equivalent to the median when $p_{*}$ is perpendicular to $p_{i}$, and then Equation \ref{eq_p_stat} reduces to $p_{*}=\sqrt{p^2-p_{i}^2}$.}
\label{fig_vector}
\end{figure}

The analysis of the PlanetPol survey data is presented here for the first time. In the PlanetPol data, debris disk systems show slightly higher median polarisations at shorter average distances. If one were to apply the $p/d$ relation established by \citet{bailey10} for stars between 10 and 17 hours RA in the PlanetPol survey, then one would expect 3.8 ppm at 19.0 pc. However as both the debris disk and main sequence samples contain stars with greater RA, a more cautious value of 5.1 ppm is achieved by scaling the $p/d$ from the other class V stars. The relation described by Figure \ref{fig_vector} then gives us a median polarisation contribution from debris disks in the PlanetPol sample of 8.1 ppm. Given PlanetPol's 5$\arcsec$ aperture wouldn't often capture the full disk for such close objects this is in line with expectations.

In the HIPPI data we have a higher polarisation for debris disks, but at a larger average distance. However $\gamma$ TrA has a tiny infrared excess and would not be expected to have much of a polarisation signal, maybe 5 ppm at most \citep{marshall15}, and yet at 56 pc it is contributing most to the mean distance. The other two disk systems together have a median polarisation of 29.1 ppm at an mean distance of 12.9 pc, which given their median excess of $F_{\rm IR}/F_{\star}=$ 18$\times$10$^{-6}$ \citep{marshall15} is more in line with expectations. However, the very small numbers of debris disk systems (and also other main sequence stars) means that it is difficult to draw conclusions at this level, particularly in light of an uncertain contribution from the local interstellar medium (Section \ref{sec_spatial}).

\begin{table}
\caption{Polarisation for debris disk and non-debris disk main sequence stars.}
\centering
\tabcolsep 3 pt\tabcolsep 3 pt
\begin{tabular}{l r r c r r c}
\hline
Type & \multicolumn{3}{c}{HIPPI$^a$} & \multicolumn{3}{c}{PlanetPol$^b$} \\
    & N       & Mean        & Median       & N     & Mean       & Median    \\
    &         & d (pc)   & p (ppm)    &       & d (pc)  & p (ppm) \\  
\hline
Debris Disk$^{c}$ &    3   &  27.4       &   29.5       &  5    &   19.0     &   9.6      \\
Other class V$^{d}$     &    6   &  16.2       &   18.1       & 11    &   33.0     &   8.8      \\
\hline
\end{tabular}
\begin{flushleft}
a - This work. \\
b - \cite{bailey10}. \\
c - If $\gamma$ TrA in HIPPI survey is excluded: 2, 12.9, 29.1. \\
d - A-K stars only. Doesn't include $\alpha$ Cen in HIPPI survey (see Section \ref{sec_alpha_cen}); if included: 7, 14.1, 27.5. \\
\end{flushleft}
\label{tab_dd}
\end{table}

Comparison of Figure \ref{fig_Equ} and Figure \ref{fig_EquPdist} reveals four of the eight debris disk systems to have polarisation angles closely aligned with their nearest neighbour stars in the survey. This naturally leads to two hypotheses: (1) that interstellar polarisation is swamping the contribution from the disk, or that (2) the local interstellar magnetic field plays a role in the formation of the system. In this instance both of these may be discounted by consideration of the individual systems. The four systems with similar alignment are Fomalhaut, $\iota$ Cen, Merak and $\gamma$ Oph. Taking the second hypothesis first, this would seem most probable if the systems were young and had formed near to where they are now. None of them is particularly young though. Furthermore, the HIPPI aperture does not encompass the whole disk of Fomalhaut nor Merak. Meaning that the position angle we have measured does not correspond to the minor axis of the disk for either of these systems \citep{marshall15}.

The contribution of the interstellar polarisation to that measured for the debris disk systems is harder to gauge without more precise data from nearby systems. Certainly, the data presented in Table \ref{tab_dd} suggest that the interstellar contribution could be significant. A number of factors argue against interstellar polarisation being dominant though. In the case of Fomalhaut the $p/d$ is 3.06 ppm~pc$^{-1}$, higher than for any other survey star identified as having interstellar polarisation alone. For $\iota$ Cen we have obtained data in different wavelength bands as part of another project (Marshall et al., unpub. data); preliminary analysis of these data gives a wavelength dependence inconsistent with interstellar polarisation \citep{serkowski73,oudmaijer01}. For $\gamma$ Oph the polarisation angle measured is very well aligned with the minor axis of the disk determined from imaging \citep{marshall15}, making the interstellar contribution very difficult to gauge. The PlanetPol measurements of Merak and $\beta$ Leo are consistent with interstellar polarisation -- $\beta$ Leo has a polarisation of only 2.3$\pm$1.1 ppm and we have, on occasion, used it as a low polarisation standard -- and it could be the characteristics of the disks in these two systems do not generate significant polarisation. For a fuller discussion of factors contributing to debris disk polarisation see \citet{marshall15}.

\subsubsection{Altair}

Altair (BS 7557, $\alpha$ Aql), observed as part of the PlanetPol survey \citep{bailey10}, though not a classical debris disk system, has been identified as having an infrared excess through interferometric measurement of the star at near infrared wavelengths \citep{absil13}. Based on the spectral slope of the excess it was predicted that the excess was attributable to scattered light rather than thermal emission. Given that small dust grains should be more effective scatterers at shorter wavelengths an appreciable polarisation would have been expected. The PlanetPol measurement is consistent with the contribution from the interstellar medium. From this we might infer that the grains responsible for scattering could well be too small to be effective polarisers, i.e. nanoscale dust grains as postulated by \citet{su13}.

\subsubsection{$\rho$ Pup}
Another non-traditional disk system is the bright giant $\rho$ Pup (HIP 39757, F5II). It was identified as having a debris disk by \citet{rhee07}, with a 60 $\mu$m excess of 5$\pm$1$\times10^{-6}$; it has an even more significant excess according to \citet{mcdonald12} of 160.7$\times10^{-6}$. The degree of polarisation measured, 18.3 ppm, is consistent with what is expected from its infrared excess \citep{marshall15}. However, there are no imaging data available to gauge its geometry or confirm that the dust present is in the form of a traditional debris disk. The system is 18.3 pc distant, positioned on the border of the low and high polarisation regions in Figure \ref{fig_EquPdist}, and so it is difficult to determine the relative contributions of the interstellar medium and any intrinsic component due to a disk.

\subsubsection{$\epsilon$ Sgr}
\label{sec_epsilon_sgr}

There is one other debris disk system in the survey not mentioned up to this point: $\epsilon$ Sgr (HIP 90185). It has a high polarisation measurement of 162.9$\pm$4.4 ppm. It is an unusual system to be discussed in the context of debris disks on two counts: (1) the primary star is a 3.52 \textit{M$_\odot$} B giant having spectral type B9.5III, and (2) it is a binary system with the 0.95 \textit{M$_\odot$} companion \citep{hurbig01} separated on a similar scale to the debris disk \citep{rodriguez12}. The secondary orbits it at 106 AU, whilst the disk has been detected in \textit{IRAS} 60 $\mu$m to have an excess of 4.5$\times$10$^{-6}$ by \citet{rhee07} and is presumed to be centred at 155 AU as a result \citep{rodriguez12}. Even greater excess has been detected with \textit{Spitzer} at 13 $\mu$m and 31 $\mu$m \citep{chen14,mittal15}, implying a closer disk. The 60 $\mu$m determination places the secondary well within the HIPPI aperture, but the disk on the edge of it. $\epsilon$ Sgr was observed on 1/9/2014 when the seeing was particularly bad ($\sim$4$\arcsec$), so we probably have a large contribution to the observed polarisation due to scattering from the debris disk.

What makes this system particularly interesting is that in V band a simple calculation shows that the contribution of the secondary to the light reaching the disk varies from $\sim$0.1\% of the total for the furthest part to $\sim$3\% for the closest. There is therefore an asymmetry in reflected light over the whole disk and thus in polarisation as well. If the polarisation we see is a result of this asymmetry then the measured polarisation angle should be related to the position angle of the binary system. The position angle of the system was measured in March 1999 to be 142.3$\degr$ \citep{hurbig01}. Our measured polarisation angle is 38.1$\pm$1.6$\degr$. When one considers that the secondary could have rotated by as much as $\sim$11$\degr$ (assuming a face-on system, a circular orbit and the formal limit on the uncertainty of the mass of component A), and that there must be contributions from interstellar polarisation, it is probable that we have measured a polarisation angle perpendicular to the position angle of the binary system. This is consistent with the stated hypothesis. If the system is inclined then the position angle of the binary system will not have swept through as great an angle, but we don't know precisely the interstellar contribution nor the contribution resulting from any asymmetry associated with disk inclination, so the small difference between the position angle and the expected polarisation angle is not significant. More significant is that the measured infrared excess is only 11.0$\times$10$^{-6}$ \citep{mittal15}, and this shouldn't be enough to generate this degree of polarisation. However, the distance of the debris disk has been determined from the \textit{IRAS} measurement at 60 $\mu$m and the shorter wavelength measurements made with \textit{Spitzer} imply either a closer disk which would produce a greater asymmetry, or a circumsecondary disk as has been suggested for the HD 142527 system \citep{rodigas14}. 

Based on the \textit{IRAS} and \textit{Spitzer} data we modelled the spectral energy distribution of the system using two blackbody components and obtained an fractional excess of 31$\times$10$^{-6}$, which is more consistent with what we would expect based on the polarisation measured. The quality of this value is strongly dependent on the assumptions made regarding the temperature of the cold component and the stellar photosphere contribution. The stellar photosphere was represented by a Castelli-Kurucz model \citep{castelli04} with an effective temperature of 10,000 K and a surface gravity, $\log g$, of 4, and solar metallicity. This was scaled to the optical and near-infrared photometry from SIMBAD. In our model we fixed the temperature of the cold component to 85 K, such that its emission peaked at 60 $\mu$m -- the wavelength of the longest reported flux density measurement. The warm component was fitted to the mid-infrared excess, which resulted in a temperature of 300 K. The combined model overestimates the flux density at 31 $\mu$m compared to that reported by \citet{chen14}. Due to the absence of longer wavelength data constraining the peak of the cold emission, the total fractional excess is subject to large uncertainties.

At this point it should be noted that $\epsilon$ Sgr B was found as a result of a search of late-B stars showing high X-ray fluxes \citep{hurbig01}. Some X-ray binaries have been found to show variable polarisation \citep{clarke10} and so this is an alternative explanation for the polarisation observed. However, such detections have been rare, and generally for much stronger X-ray sources. Considering the magnitude of the flux asymmetry induced by the secondary, scattering from the debris disk seems the most likely mechanism. Indeed the X-ray activity might be an indication of dust accretion onto the secondary, which is another scenario suggested for HD 142527 \citep{rodigas14}.

Presuming the polarisation we see in this system is a result of scattering from the disk, then the behaviour with wavelength will be a function of the ratio from components A and B, as well as the properties of the grains in the disk. The polarisation will likely be greater at redder wavelengths than typical for a debris disk system. If the secondary is interior to the disk, as it orbits it will race ahead of the disk, illuminating each section in succession as if a torch shinning upon it. Follow up observations of this object offer a unique opportunity to use polarisation to probe the radial homogeneity of the debris disk. Though it will take some time...

\subsection{Ap stars}
\label{sec_ap}

As an aside, from Figure \ref{fig_EquPdist} it is clear that $\epsilon$ UMa (BS 4905) is more polarised than the stars around it. $\epsilon$ UMa has spectral type A1III-IVp and is the brightest chemically peculiar star in the sky. Ap stars typically exhibit high degrees of polarisation as a result of magnetic fields in the many hundreds of Gauss \citep{clarke10}. Such fields generating broadband polarisation through, for example, differential saturation of Zeeman components as described by \citet{leroy90}. $\epsilon$ UMa has a strong magnetic field, and a rotational axis at an angle to its magnetic axis that periodically brings its magnetic pole into line of sight; this is the likely cause of its high degree of polarisation. However, recently it was proposed that the periodicity displayed (5.0887 d) by the star is a consequence of a 14.7 M$_J$ companion orbiting at a distance of 0.055 AU \citep{sokolov08}, and this could also be a contributor. 

There is one other Ap star in the PlanetPol survey: $\alpha^2$ CVn. This star is known to have a regular time varying polarisation. Recent broadband measurements in a similar range to that of PlanetPol have shown that the degree of linear polarisation varies between zero and more than 0.7\% \citep{kochukhov10}. It thus appears a matter of chance that the recorded degree of polarisation for $\alpha^2$ CVn by PlanetPol was just 8.8 ppm.

\subsection{Eclipsing binaries}
\label{sec_eclipse}

The work credited with bringing about the beginning of stellar polarimetry in earnest is that of \citet{chandrasekhar46}\footnote{The serendipitous discovery of interstellar polarisation was made in the process of searching for the \textit{Chandrasekhar Effect} \citep{clarke10}.}. In this work he described how polarisation arising from free electron scattering at the limb of a star might be observed by using an eclipsing binary to break the symmetry of the stellar disc. It was shown that the limb would be most polarised with the azimuth of vibrations tangential to the limb, and that the polarisation would fall away quickly from there, becoming zero in the centre of the stellar disc. Usually the symmetry of the stellar disc renders this effect undetectable in aperture polarimetry, but an eclipsing binary breaks the symmetry. The magnitude of the limb polarisation varies depending on the type of star, with earlier types showing greater polarisation \citep{kostogyrz15b}. Measurements of polarisation across the disc of the Sun were first used to confirm the effect, and the Sun has a maximum limb polarisation of $\sim$12\% \citep{kostogyrz15b}. 

There is one eclipsing binary in our survey, $\delta$ Cap (HIP 107556, A7III). This star also happens to be chemically peculiar. The G type secondary has an orbital period of 1.022789 days \citep{eggleton08}. It is noteworthy that the error associated with our measurement of $\delta$ Cap is larger than any other in the survey. This could be the result of observing the system during a transit or transitioning into or out of secondary eclipse producing variable limb polarisation effects. However, the weather was variable on the night of the observation, and it is equally likely the large error is associated with reduced signal from patchy cloud.

The star $\alpha$ CrB (BS 5793, A0V) observed in the PlanetPol survey is also an eclipsing binary of the Algol type \citep{eggleton08}, but doesn't show any polarimetric behaviour from those observations of note.

\subsection{Close binaries}
\label{sec_binary}

Variable polarisation can occur in close binary systems as a result of light scattered from material co-rotating in the system or from emission line effects associated with stellar winds or gaseous streams \citep{clarke10}. A common property of many early-type close binary systems is the presence of a gaseous extrastellar envelope, possibly of proto-solar material \citep{mclean80}. The dynamics of a binary system render this envelope asymmetric, resulting in an intrinsic polarisation signal that varies with the binary phase according to the system geometry with respect to the observer and the polarigenic mechanism \citep{mclean80}. The requirement for extrastellar material resulted in a divide between evolved and unevolved stars when investigated by \citet{pfeiffer77}. Having investigated systems with separations up to $\sim$5 AU, they found binaries with unevolved stars, with few exceptions, do not show intrinsic polarisation; whilst those with evolved stars are likely to be intrinsically polarised if the pair are separated by more than 10 solar radii (10 R$_\odot$). \citet{pfeiffer77} hypothesised that the 10 R$_\odot$ divide resulted from insufficient material being present to generate a large enough signal to be detectable by the instrumentation of the time (that being $\sim$10$^{-4}$).

Within our survey there are a three B-type primaries with close companions, these being HIP 60718 ($\alpha$ Cru -- as already mentioned in Section \ref{sec_alpha_cru}), HIP 100751 ($\alpha$ Pav) and HIP 65474 (Spica). All three show significant polarisations.

\subsubsection{Spica}
\label{sec_spica}

Spica is classified as B1III-IV and has a B2V companion with an orbital period of just 4.0145898 days \citep{harrington09}. It is Beta Cephei-type variable star that varies in brightness with a 0.1738-day period as a result of its outer layers pulsing. The brightness variability has been shown to be due to line-profile variability as a consequence of surface flows induced by tidal forces \citep{harrington09}.

Spica was investigated by \citet{pfeiffer77} and found not to vary polarimetrically -- having an observed polarisation of 0.03$\pm$0.01\% it was designated as being unpolarised. However, as our initial measurements of Spica were intermediate of those of \citet{tinbergen82} given in Table \ref{tab_previous} -- even if nominally within the error in $q$ -- we considered it an interesting target to follow up. Table \ref{tab_spica} shows the three measurements we have made of Spica. From these it is clear that it is varying and thus intrinsically polarised.

\begin{table}
\caption{Individual measurements of Spica.}
\tabcolsep 3 pt
\centering
\begin{tabular}{llrrrr}
\hline
Date$^a$ & UT$^b$ &
       \multicolumn{1}{c}{q (ppm)} & 
       \multicolumn{1}{c}{u (ppm)} & 
       \multicolumn{1}{c}{p (ppm)} &
       \multicolumn{1}{c}{$\theta$ $(\degr)$} \\  \hline

24/5 & 11:02 & -154.0 $\pm$ 2.5 & 24.7 $\pm$ 2.5 & 156.0 $\pm$ 2.5 & 85.4 $\pm$ 0.9 \\
29/6 & 10:32 & -185.1 $\pm$ 2.9 & 30.6 $\pm$ 2.8 & 187.6 $\pm$ 2.9 & 85.3 $\pm$ 0.9 \\
29/6 & 12:49 & -188.1 $\pm$ 3.6 & 57.2 $\pm$ 3.9 & 196.6 $\pm$ 3.8 & 81.6 $\pm$ 1.1 \\
\hline
\end{tabular}
\begin{flushleft}
a - All dates are 2015. \\
b - The time given as hh:mm and is that corresponding to the beginning of the measurement from the third telescope position angle in the sequence. The target was reacquired for each position angle leading to some minor variation in the timing. There was also a short pause between the middle two measurements of the third observation owing to passing cloud. Each measurement represents 24 mins of data plus acquisition time. \\
\end{flushleft}
\label{tab_spica}
\end{table} 

\subsection{Be stars}
\label{sec_be_stars}

Be stars are defined as non-supergiants of B spectral type that have exhibited episodic Balmer line emission. The origin of the emission is attributed to the ejection of a gas circumstellar envelope (CSE) by the star \citep{domiciano03}. Be stars are rapid rotators that exhibit episodic mass and angular momentum losses as well as disc formation and dissipation \citep{domiciano14}. The rotation rate is usually said to be 70-80\% of the star's critical velocity \citep{porter03}. The stellar winds of Be stars are faster than those of ordinary B stars, particularly in earlier types. The winds are asymmetric, characterised by fast tenuous winds at the poles and stronger slower winds at equatorial latitudes. 

Such stars exhibit varying polarisation on both short and long timescales. On minute to hour timescales fluctuating polarisation is attributed to ejection events, where a `blob' of material is formed at the stellar surface, distorting its shape, which co-rotates with the star with the Keplerian velocity of the inner decretion disc \citep{carciofi07}. The extra equatorial material adds to the already distorted shape of the star caused by its sub-critical rotation (which is also responsible for gravity darkening at equatorial latitudes) \citep{domiciano03}. On longer time scales a circumstellar decretion disc of ionized gas is built up around the star, the presence of which also produces an infrared excess and a non-zero (usually dominant) polarisation signal according to its density and geometry -- the polarisation angle is aligned with the rotational axis of the star \citep{carciofi07}. During a quiescent phase the disc is dissipated by radiation pressure and partial reaccretion onto the star \citep{carciofi12}. Be stars are known to be variable on time scales ranging up to decades. A quiescent Be star might appear as an ordinary B star before becoming active within just a couple of days \citep{peters86, barnsley13}. A subclass of these objects is the B$_{shell}$ star, where the emission lines are depressed as a result of viewing geometry, i.e. they star is equator-on to the observer \citep{saad12}. However, B$_{shell}$ behaviour is often seen intermediate of that of B and Be behaviour, suggesting the CSE may sometimes develop, at least initially, with an imperfect alignment. In such systems polarimetry is a relatively more sensitive probe of disc development.

The definitive statistical study of Be stars and their polarisation is that of \citet{yudin01}; Table \ref{tab_be} is a reproduction of the data he collated for the Be stars in our survey. 

\begin{table}
\caption{Observed polarisation for Be stars from \citet{yudin01} compared to our measurements.}
\tabcolsep 3 pt
\centering
\begin{tabular}{lcccc}
\hline
Star & \multicolumn{2}{c}{\citet{yudin01}$^a$} & \multicolumn{2}{c}{This Work} \\
HIP        & $p$ (\%)   & $\theta$ ($\degr$) & $p$ (\%)   & $\theta$ ($\degr$)       \\
\hline
7588    & 0.04 \hspace{7.5 mm}  & 136 & 0.21518 $\pm$ 0.00038$^b$ &   \hspace{1.5mm}31.2 $\pm$ 0.1 \\ 
26634   & 0.15 $\pm$ 0.05 & 109 & 0.07220 $\pm$ 0.00064   &   101.1 $\pm$ 0.5   \\ 
71352   & 0.61 $\pm$ 0.04 & 174 & 0.53206 $\pm$ 0.00038   &   172.0 $\pm$ 0.0   \\ 
85792   & 0.37 $\pm$ 0.07 & 174 & 0.62400 $\pm$ 0.00034   &   171.8 $\pm$ 0.0   \\ 
\hline
\end{tabular}
\begin{flushleft}
a - Note that the tabulated values are the observed values, and that for some objects \citet{yudin01} also calculated the intrinsic and interstellar components. \\
b - Tabulated value for this work is the average of two measurements. \\
\end{flushleft}
\label{tab_be}
\end{table}

Our measurements of $\alpha$ Col (HIP 26634), $\alpha$ Ara (HIP 85792) and $\eta$ Cen (HIP 71352) in the $g^{\rm \prime}$ agree in position angle with those collated by \citet{yudin01}. For $\alpha$ Ara \citet{meilland07} resolved the equatorial disc using VLTI/AMBER data and calculated models with orientation in agreement with the position angle given by \citet{yudin01}, after subtraction of an interstellar component of 0.15\% at 30$\degr$ to give 166$\degr$. Given the discussion in Section \ref{sec_spatial}, this determination of the degree of interstellar polarisation seems high. \citet{meilland12} conducted a spectro-interferometric survey of Be stars which included both $\alpha$ Ara and $\alpha$ Col. The position angle for the rotational axis of these stars was obtained by means of a axisymmetric kinematic model. The determined position angles were 88$\pm$2$\degr$ and 10$\degr$ for $\alpha$ Ara and $\alpha$ Col respectively. This places our determined position angle for the circumstellar disc of $\alpha$ Ara between that of Meilland et al.'s 2007 and 2012 determinations, and our measurement for $\alpha$ Col in closer agreement with \citet{meilland12} than that of \citet{yudin01} is.

Our measurement for the degree of polarisation of $\alpha$ Col is half the tabulated value indicating a smaller circumstellar disc, whereas for $\eta$ Cen we record almost double the polarisation level, indicating a thicker circumstellar disc. For $\alpha$ Ara we record a similar level of polarisation. There is no agreement for Achernar (HIP 7588) though, and as we observed it twice it bears closer scrutiny.

\subsubsection{Achernar}

As one of the closest brightest Be stars Achernar has been extensively studied from the 1970s onward. During that time it has been in and out of emission -- gaining and losing its disc. During the most recent period quiescent period \citet{desouza14} took advantage of a negligible disc in using interferometry to measure the position angle of Achernar's rotational axis as 216.9$\pm$0.4$\degr$. They also measured the rotational flattening of Achernar as $R_{eq}/R_{p}$ = 1.352$\pm$0.260 making it one of the flattest fast-rotating stars. When the star is in an emission or shell state it will have a significant disc that would be expected to lie with its major axis perpendicular to the star's rotational axis. 36.9$\pm$0.4$\degr$ is thus the angle one would expect from a polarimetric measurement of the star in a disc bearing phase before any other considerations. Historic measurements of the polarisation of Achernar together with our new measurements are given in Table \ref{tab_achernar}.

\begin{table*}
\caption{The polarisation of Achernar over time.}
\centering
\begin{tabular}{lllllllcl}
\hline
Date        &  \multicolumn{3}{c}{$p$ (\%)}  &  \multicolumn{3}{c}{$\theta$ ($\degr$)} & Band & Reference\\
\hline
31/1/1968       &   0.03    &$\pm$&   0.01    &   \hspace{1.5 mm}26      &$\pm$&   10      &   V   &   \citet{serkowski70} \\
12/12/1968      &   0.02    &$\pm$&   0.01    &   \hspace{1.5 mm}46      &$\pm$&   14      &   V   &   \citet{serkowski70} \\
1969-70$^a$     &   0.011   &$\pm$&   0.004   &   136.2   &$\pm$&   10.4    &   V   &   \citet{shroeder76}  \\
1977-78$^a$     &   0.001   &$\pm$&   0.005   &           &     &           &   400-700 nm & \citet{tinbergen79} \\
1/9/1995$^b$    &   0.14    &$\pm$&   0.04    &   \hspace{1.5 mm}41      &$\pm$&   \hspace{1.5 mm}8       &   V   &   \citet{mcdavid05}  \\
1/9/1995$^b$    &   0.11    &$\pm$&   0.05    &   \hspace{1.5 mm}45      &$\pm$&   13      &   B   &   \citet{mcdavid05}  \\
7/2006$^c$      &   0.159   &$\pm$&   0.003   &   \hspace{1.5 mm}31.7    &$\pm$&   \hspace{1.5 mm}0.6     &   V   &   \citet{carciofi07} \\
9-11/2006$^c$   &   0.130   &$\pm$&   0.001   &   \hspace{1.5 mm}32.6    &$\pm$&   \hspace{1.5 mm}0.3     &   V   &   \citet{carciofi07} \\
9-11/2009$^{cd}$&   0.011   &$\pm$&   0.013   &           &     &           &   B   &   \citet{desouza14}  \\
29/6/2011$^d$   &   0.016   &$\pm$&   0.038   &           &     &           &   B   &   \citet{desouza14}  \\
9-10/2011$^{cd}$&   0.020   &$\pm$&   0.015   &           &     &           &   B   &   \citet{desouza14}  \\
9-11/2011$^{cd}$&   0.015   &$\pm$&   0.009   &           &     &           &   V   &   \citet{desouza14}  \\
1/7/2012$^d$    &   0.017   &$\pm$&   0.015   &           &     &           &   V   &   \citet{desouza14}  \\
21/11/2012$^d$  &   0.035   &$\pm$&   0.050   &           &     &           &   V   &   \citet{desouza14}  \\
2/9/2014        &   0.26984 &$\pm$&   0.00045 &   \hspace{1.5 mm}31.5    &$\pm$&   \hspace{1.5 mm}0.1     &   g$^{\rm \prime}$  &   This work.         \\
24/5/2015       &   0.16052 &$\pm$&   0.00062 &   \hspace{1.5 mm}31.8    &$\pm$&   \hspace{1.5 mm}0.2     &   g$^{\rm \prime}$  &   This work.         \\
\hline
\end{tabular}
\begin{flushleft}
a - Exact dates of observations not reported. \\
b - Measurements for U, R and I bands also reported in \citet{mcdavid05} are the same within error as the B and V band measurements. \\
c - For the sake of brevity we have averaged similar measurements reported by \citet{carciofi07} and \citet{desouza14}. In the case of the former the measurements were weighted by the number of wave-plate positions indicated. Where p/$\sigma$ $>$ 4 debiasing has been carried out as $\sqrt{p^2-\sigma^2}$, otherwise the median is given for \textit{p}, as recommended by \citet{clarke10}. \\
d - Position angles not reported. \\
\end{flushleft}
\label{tab_achernar}
\end{table*}

\citet{mcdavid05} interpreted the results of \citet{shroeder76} as being consistent with measuring the rotationally oblate photosphere of Achernar during a phase where the circumstellar disc was completely absent. However, the degree of polarisation recorded, even allowing for uncertainty, is more than double what one would expect from the calculations of \citet{sonneborn82} given Achernar's established inclination angle of $\sim$60$\degr$ \citep{desouza14} to $\sim$65$\degr$ \citep{carciofi07}. Alternative explanations are that this is the interstellar component of the observed polarisation, or that in the absence of the CSE it might be due to Achernar's polar wind \citep{stee09} or a combination of one or more of these.

Excepting that of \citet{shroeder76} the data in Table \ref{tab_achernar} fall into two distinct categories: those where there was a significant disc present, and those where there was, perhaps, only a tenuous disc. The distinction being approximately an order of magnitude in \textit{p}. The earlier position angle determinations are consistent with that expected from the interferometry measurements made by \citet{desouza14}. However, our measurements, and the averages of those made by \citet{carciofi07} are more precise and disagree by $\sim$5$\degr$. \citet{carciofi07} observed the polarisation to change in both degree and angle on a timescales of days to weeks, as well as on hour long timescales. They interpreted the change in position angle as likely associated with a departure from a simple 2D geometry for the disc. Changes in position angle of the order of 5$\degr$, and in the degree of polarisation of $\sim$0.02\% were seen to occur on timescales of less than an hour. 

Here our values represent observations made over a little less than half an hour each, so it's reasonable to assume we've seen a distorted disc geometry due to the addition of transitory blobs. Our measurement of 2nd September 2014 represents the highest polarisation ever measured for Achernar (the previous highest being that of \citet{carciofi07} on 14th July 2006 of 0.188$\pm$0.009\%) and thus the greatest recorded build-up of circumstellar material. By the 24th of May 2015 the polarisation level had dropped by a third, but was still, historically speaking, very high.

From Table \ref{tab_achernar} our measurements are far more precise than those recorded previously. HIPPI therefore has a previously unseen capacity to probe the growth and dynamics of circumstellar discs around Be stars. The growth and dynamics of circumstellar discs are still poorly understood \citep{carciofi07}, high precision polarimeters like HIPPI clearly offer an enhanced capacity to study this phenomenon. 

\subsection{B stars}
\label{sec_b_stars}

Setting aside the Be stars Table \ref{tab_spc} makes it clear that B stars are more polarised than A, F, G, and K type stars. The starkness of the transition from A stars is shown up particularly well in Figure \ref{fig_HR}; it also appears that early B stars are more polarised than later B stars. After excluding the Be stars, the close binaries and the unusual debris disk system $\epsilon$ Sgr, there are five stars of B spectral type remaining. Although the case for $\gamma$ Crv and $\alpha$ Mus (which is 96.7 pc distant) is marginal, they all have polarisations that are higher than would be expected from only interstellar polarisation when considering their positions in Figures \ref{fig_EquPdist} and \ref{fig_Pwd}. 

Early type stars can be expected to have higher magnetic fields in general as a consequence of their younger age. Some Bp \citep{briquet14} and Be stars \citep{hurbig13} have been shown to have significant magnetic fields (though the fields in Be stars are weak -- less than $\sim$100 Gauss). However, no magnetic fields have been detected in `normal' B stars within a margin of error of 13 Gauss \citep{shorlin02}. Consequently we don't expect these stars to show polarisation from differential saturation of Zeeman components or similar mechanisms.

One possibility to account for the high degree of polarisation is that these might be debris disk host stars. Infrared excess associated with debris disks is preferentially found in young systems. However, \citet{padgett12} have recently, from \textit{WISE} data, given the portion of B stars within 120 pc hosting debris disks as 12.1$\pm$2.0\%. Similarly \citet{schultz75} found an infrared excess -- indicative of a debris disk system -- in 20 of 350 O and B stars. Given that we have already identified one such system in our survey, the probability of many more is low.

A feature of B type stars is that many of them have a high rotational velocity. From Table \ref{tab_stars} it can be seen that $\alpha$ Gru, $\beta$ Lib and $\sigma$ Sgr all have quite high rotational velocities. The spread in distances of these objects is quite large, and so we've plotted $p/d$ against rotational velocity in Figure \ref{fig_b_stars} to attempt to reduce the effects of interstellar polarisation. Figure \ref{fig_b_stars} shows B stars with higher rotational velocities tend to have higher polarisations. It appears that there is an exponential increase in $p/d$ with $V$~sin~$i$. There are no exceptions to this trend in our data, but there are two stars surveyed by PlanetPol that don't fit.

$\tau$ Her has a high $p/d$, especially amongst stars from the PlanetPol survey. However, it lies at 94.6 pc distance, where we have very few stars, and the polarisation of the interstellar medium is not well constrained. The degree of polarisation we measured could be indicative of the edge of the Local Bubble -- as originally suggested for this object by \citet{bailey10}. So, the polarisation measured could be interstellar in origin, but the lack of other objects at similar distances in the same part of the sky makes it hard to draw conclusions. The polarisation of $\delta$ Cyg is not so easily explained away as being interstellar. It's a multiple system but the B component (F1V) is sufficiently separated that we would not expect this to be the polarigenic mechanism in this case. It could be that $\delta$ Cyg has a high rotational velocity, but a low inclination giving a low value for $V$~sin~$i$, but this is purely speculation.

There are two candidates for a polarigenic mechanism related to rotational velocity. We begin by investigating that proposed by \citet{bailey10} for Regulus (BS 3982) -- the star with the highest $V$~sin~$i$ in Figure \ref{fig_b_stars}. 

\begin{figure}
\includegraphics[width=0.5\textwidth,natwidth=100,natheight=100,trim=0.5cm 0.5cm 0.5cm 0.5cm]{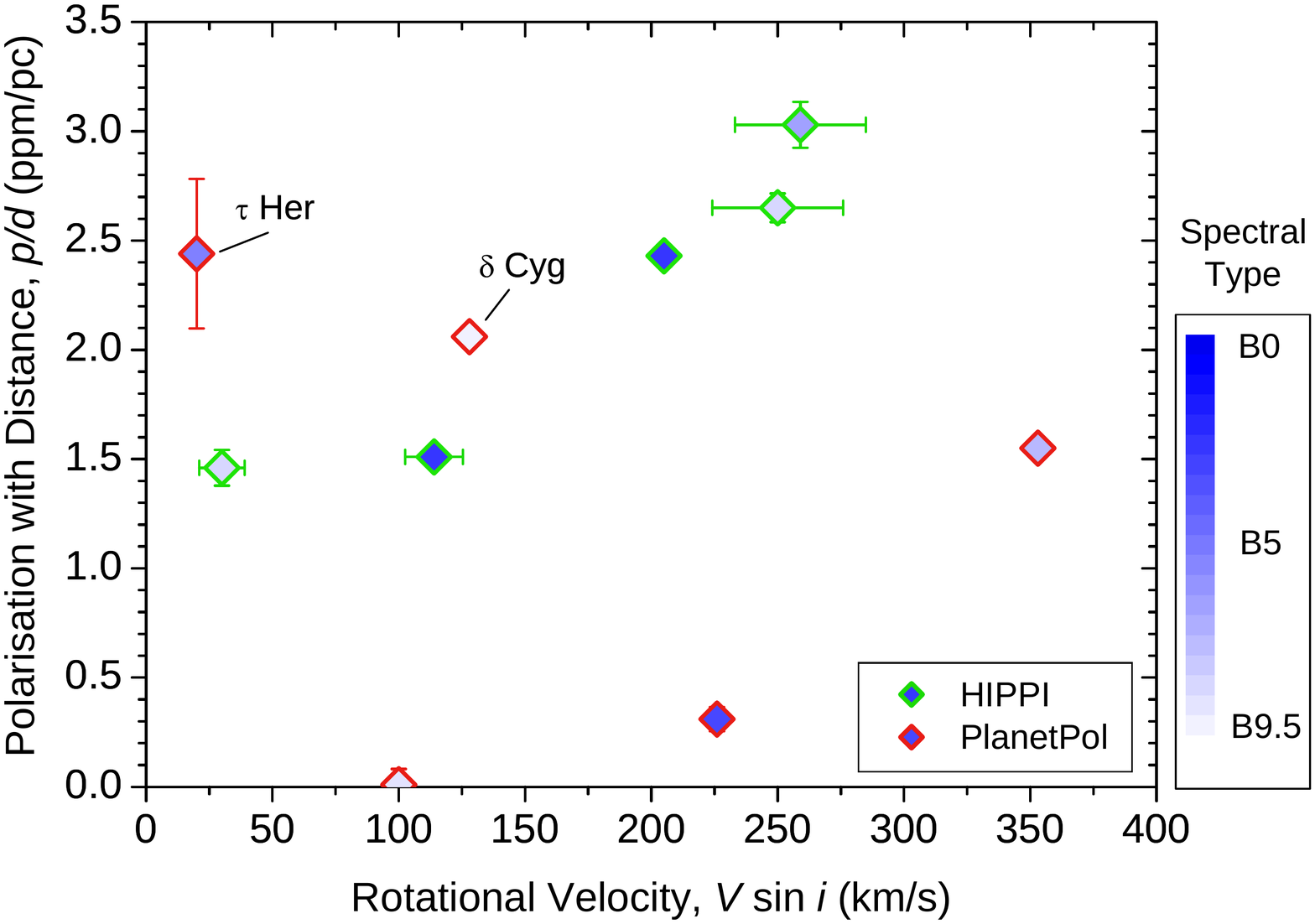}
\caption{Polarisation with distance, $p/d$ as a function of rotational velocity for B stars without other identified polarigenic mechanisms. For the stars in our survey there is a clear trend in increased polarisation with rotational velocity, whilst for the stars in the PlanetPol survey two objects don't fit the trend: $\tau$ Her and $\delta$ Cyg. Parameters for the PlanetPol stars come from \citet{bailey10}.}
\label{fig_b_stars}
\end{figure}

\subsubsection{Rotational oblateness}
\label{sec_fast_rotators}

Based on the work of \citet{chandrasekhar46} on revealing limb polarisation with an eclipsing binary used to break the symmetry of the disc, \citet{ohman46} reasoned that a rotationally flattened star must show polarisation at all times. Detailed modelling of the expected behaviour has since quantified the effect expected, with the most recent predictions relevant to broad band polarimetry being made by \citet{sonneborn82}. The magnitude of the polarisation expected is now believed to be much lower than originally suggested, and in almost 70 years of searching, polarisation induced by electron scattering in a rotationally flattened atmosphere has not yet been confirmed.

A degree of polarisation of 500 ppm parallel to the rotational axis is expected at 450 nm for a B0 star inclined at 90$\degr$ and rotating at 95\% of its critical velocity. This amount is greatly reduced for inclination angles more face on. It also decreases with increasing spectral type, with the maximum in the visual also shifting redder. The progression is dramatic with a B5 star expected to show at most 50 ppm at 700 nm \citep{sonneborn82}.

As already mentioned, \citet{bailey10} have proposed this mechanism for the B6 star Regulus, which rotates at 86\% of its critical velocity \citep{mcalister05}. Regulus is close by in a part of the sky with low interstellar polarisation and registered 36.8$\pm$1.6 ppm at an effective wavelength of 741.2 nm with PlanetPol. We believe Regulus still to be the best candidate for demonstrating polarisation due to rotational oblateness, but unless the effect has been significantly underestimated it cannot explain the high polarisation of the other B-type stars seen here.

We see larger polarisations in $g^{\rm \prime}$ than were seen by PlanetPol in its redder pass-band ($\sim$575-1025 nm). There is no trend with spectral type in Figure \ref{fig_b_stars}, and our stars are rotating much slower than Regulus. Even accounting for a more significant interstellar contribution to the observed polarisation, the excess observed is far too large to be attributed to rotational oblateness.  

\subsubsection{B stars as Be stars}
\label{sec_b_as_be}

Another possibility is that the B stars that show significant polarisation are actually Be stars outside of an emission phase, or that the polarisation we see is the result of a kind of sub-Be behaviour, or perhaps evidence for a polar wind in the absence of a CSE. In particular if a small amount of ejected gas is present at the equatorial regions this could produce the observed polarisation. This hypothesis is advanced on account of the rotational velocity trend (Figure \ref{fig_b_stars}) as well as the sharp division between A and B type stars in the H-R diagram (Figure \ref{fig_HR}). The statistical analysis of \citet{yudin01} shows a triangular distribution for Be star polarisation with $V$~sin~$i$ that peaks at 200-250 km/s, we have far fewer B stars, but the distribution is not inconsistent with that picture. It has been suggested that faster rotators are more likely to have polar winds and as a result a circumstellar disc \citep{stee09}. 

A general description of Be behaviour and the transition from B to B$_{shell}$ to Be has already been given in Section \ref{sec_be_stars}. In early type Be stars the strong winds are enough to drive material away from the star creating the CSE. That the winds of B stars are weaker might be a problem for this hypothesis, but later type Be stars have weaker winds and still show Be behaviour. Other mechanisms like non-radial pulsations, magnetic activity and binarity have been advanced to explain CSE development in later Be stars \citep{saad12,stee09}.

It is worth keeping in mind that the levels of polarisation we are seeking to explain are only $\sim$100 ppm -- the same level recorded by \citet{shroeder76} for Achernar attributed to a disc-free state. This is more than three orders of magnitude lower than the level modelled by \citet{wood97} for the Be star $\eta$ Tau for instance. Naively scaling the amount of infrared excess they modelled for their thin disc model in correspondence with the degree of polarisation we see here would result in something very difficult to detect given that infrared excess is attributable to gas \textit{and} dust. Indeed from Table \ref{tab_stars} it can be seen that the excess from B stars is not too different from that of Be stars in the survey. So, it may well be that polarisation is now a more sensitive probe of close circumstellar gas than infrared excess. It is certainly more sensitive than emission line measurements, as when there is little gas there is no emission. Furthermore polarisation is more sensitive to the inner part of any disc, whereas infrared is a better probe of the outer parts \citep{wood97}. We could be seeing a tenuous proto-CSE, and while a weak wind might not prevent a gas disc forming, it would likely prevent it developing as readily.

Recently the issue of transition from B to Be star was looked into by \citet{barnsley13}. They examined H$\alpha$ variability in a representative sample of Be stars. They formed a number of conclusions: (1) that the full phase of transition between B and Be star probably occurs over centuries, (2) that stars with earlier spectral type and higher $V$~sin~$i$ show greater variability in H$\alpha$ emission, and (3) that for stars with smaller $V$~sin~$i$ or later spectral type variability was more likely on longer (years) timescales. What we take away from this is that stars with a higher rotational velocity or an earlier spectral type are more likely to show the emission characteristics of classical Be stars. But that any B star might become a Be star at any point in time, and that for slower later stars the behaviour is more subtle.

It stands to reason then that B stars with higher rotational velocities are also more likely to form the gas shells needed for emission behaviour, even if insufficient gas is actually stockpiled to generate the emission. If this is the case we'd expect these shells to form at equatorial latitudes and result in polarisation perpendicular to the rotational axis just like in Be stars \cite{porter03,saad12,desouza14}. Furthermore, whilst variations in degree of polarisation are expected, variations in the polarisation angle should be small -- as they are with Achernar for instance. This hypothesis has the advantage of explaining the trend we see with $V$~sin~$i$ but also allowing for the exceptions of $\tau$ Her and $\delta$ Cyg. 

A pair of observations of $\sigma$ Sgr (HIP 92855, B2V) shows the expected Be-like behaviour, albeit at a marginal level (Table \ref{tab_sigma_sgr}). Our observations therefore represent evidence for sub-Be behaviour in `ordinary' B stars. Incidentally, the inclination of $\sigma$ Sgr has been estimated as 42$\pm$4$\degr$, putting its rotational velocity at 60-78\% of critical \citep{hutchings79}\footnote{Note the star name is incorrectly given as 6 Sgr in that work.} -- in line with typical Be stars. Higher than normal X-ray emission for $\sigma$ Sgr has also been seen by \textit{XMM Newton} \citep{oskinova12}, which according to \citet{hurbig13} is a behaviour associated with the Be phenomenon. 

\begin{table}
\caption{Repeat observations of $\sigma$ Sgr.}
\centering
\begin{tabular}{lcc}
\hline
Date        &  \multicolumn{1}{c}{p (ppm)}  &  \multicolumn{1}{c}{PA ($\degr$)} \\
\hline
1/9/2014        &   175.1 $\pm$ 3.8 &   130.5 $\pm$ 1.3 \\
22/5/2015       &   164.5 $\pm$ 5.9 &   129.0 $\pm$ 2.1 \\
\hline
\end{tabular}
\label{tab_sigma_sgr}
\end{table}

After $\sigma$ Sgr the next most polarised `ordinary' B star in the survey is $\beta$ Lib (HIP 74785, B8Vn). Notably it has the highest infrared excess amongst these stars (Table \ref{tab_stars}). The inclination of $\beta$ Lib has been estimated as 59$^{+11}_{-8}$$\degr$ (i.e. 45-64\% of critical velocity) by \citet{hutchings79}.  The case for $\beta$ Lib being a Be star is strengthened by its listing in SIMBAD as a variable star -- rare/occasional low-level photometric variability being associated with fading in late Be stars \citep{hubert98} -- if it is variable it must be so on long timescales as it has been recommended as a photometric standard by a variety of later sources (e.g. \citet{adelman01}, references within \citet{baade89}). Additionally \citet{mclaughlin32} made a rather extraordinary aside in examining plates of $\beta$ Lib from 1912 to 1915 (to test a claim that it was a spectroscopic binary\footnote{Even today it is uncertain whether $\beta$ Lib is a binary -- \cite{roberts07} have identified a second star $\sim$2.1$\arcsec$ away, but they indicate follow-up observations are needed to confirm it isn't a background star. If it is a companion its spectral type is estimated as M2V.}), saying, ``Only the Balmer series of hydrogen is at once evident, and these lines are so broad and diffuse that measurements are difficult, even on over exposed plates. On some spectrograms H$_{\gamma}$ and H$_{\delta}$ appear triple, \textit{as if weak emission were present}.''\footnote{This statement gains greater significance when one realises that McLaughlin authored two other papers on Be star spectroscopy in the same year \citep{mclaughlin32b,curtiss32}, and was thus well practised in identifying emission characteristics in the photographic plates of the time.} 

\subsection{Late giants}
\label{sec_giants}

The survey also contains a significant fraction of late-type giants. Two of these have significant previously measured polarisations: HIP 89931 ($\delta$ Sgr, K3III) at the 10$^{-4}$ level \citep{tinbergen82} and the SRB type pulsating variable HIP 112122 ($\beta$ Gru, M5III) also at the 10$^{-4}$ level \citep{heiles00}. Both of these stars also show polarisations at that level here. As discussed earlier in Section \ref{sec_prev} our measurements are significantly different to the earlier ones, indicating that the polarisation is variable and therefore intrinsic to the star. In addition HIP 79593 ($\delta$ Oph) has a similarly high polarisation. At 61.3 pc distance a polarisation of 131.1 ppm for HIP 110130 ($\alpha$ Tuc, K3III) might suggest that it is also intrinsically polarised. When one considers that in the PlanetPol survey that the M0 giants $\alpha$ Vul (BS 7405) and $\gamma$ Sge (BS 7635) recorded 1321.4 ppm and 199.5 ppm respectively, this amounts to four out of eight luminosity class III M giants from the two surveys that are probably intrinsically polarised, and very probably two K type class III giants that are as well, based purely on the degree of polarisation. In addition, careful examination of the spatial distribution and $p/d$ relations in Section \ref{sec_spatial} suggests that $\gamma$ Dra (BS 6705, K5III) and $\kappa$ Lyr (BS 6872, K2III) may also belong to this group. Likewise $\alpha$ Phe (BS 2081, K0.5IIIb) has a polarisation/distance of 3.17 ppm/pc, higher than any other star we consider to show interstellar polarisation only.

From Figure \ref{fig_HR} we see that the degree of polarisation tends to increase with spectral type and also with luminosity in these stars. \citet{dyck71} examined polarisation in 55 K and M giants and supergiants. They found that almost all supergiants displayed significant polarisation and that class III giants displayed an increasing likelihood of being polarised with spectral type from M3 later. However, the sensitivity of their study was no better than 10$^{-3}$ to 10$^{-4}$. Given the results presented here, it now appears that many warmer giants are also intrinsically polarised, just to lower levels. Indeed, the study by \citet{dyck71} includes $\delta$ Oph, $\alpha$ Vul and $\gamma$ Sge determining no intrinsic polarisation in each case\footnote{\citet{dyck71} also measured another three stars from the PlanetPol survey without significant polarisations.}.

\citet{dyck71} studied the wavelength dependence of the polarisation in these stars, finding that it increased as $\lambda^{-1}$ before tailing off at blue or violet wavelengths, where the exact peak varied from star to star. For an M0 spectral type the effective wavelength for HIPPI is 2.094 $\mu$m$^{-1}$ whereas for PlanetPol it was 1.264 $\mu$m$^{-1}$. This means that HIPPI is 1.66 times as sensitive to this phenomenon as PlanetPol was. Indeed, if one applies this factor to the median polarisation given for M stars by PlanetPol in Table \ref{tab_spc}, the value is closer to that obtained by our survey.

\subsubsection{Circumstellar dust}
\label{sec_giant_dust}

\citet{dyck71} also noted that all the stars in their survey with significant polarisations were variable stars and that they all had an infrared excess at 11 $\mu$m  (which is associated with TiO dust, TiO being one of the first molecules formed in a cooling stellar environment). Some variability was common, and was, in fact, one of the criteria used to assign polarisation as intrinsic. The combination of these properties resulted in the cause of the polarisation to be attributed as scattering or absorption from a circumstellar shell of solid condensates \citep{dyck71,jennings72}. In this picture, dust forms in the circumstellar environment around giant stars through a two-step process: pulsations in the outer envelope elevate molecular gas to the surface, whence the stellar wind blows it radially outward to condense as grains in a cooler environment \citep{morris87}. Dust is associated with IR excess and many red giants have significant IR excess (c.f \cite{mcdonald12}). For polarisation to be induced from the dust requires that the circumstellar envelope be asymmetric with respect to the observer. There are a number of potential causes of the asymmetry which are debated. It has been shown that the wavelength dependence of polarisation due to circumstellar dust depends on particle size and composition \citep{mccall80, raveendran91} as well as the geometry of the system. Silicate or carbonaceous grains are most common, however polarisation may also result from scattering from inner shells of condensates able to form in hotter regions such as corundum and iron-poor silicates which are transparent at NIR wavelengths \citep{ireland05}.

In all, there are 24 giants, a bright giant and a supergiant later than G5 in both surveys. Of these 18 appear in the infrared excess catalogue of \citet{mcdonald12}. All but two have an excess greater than 1$\times$10$^{-6}$ in fractional luminosity, with a range encompassing 11 $\mu$m, peaking around 12 $\mu$m or shorter. We plot polarisation with infrared excess in Figure \ref{fig_FracLwP}. This reveals that the degree of polarisation is not well correlated with infrared excess. This does not necessarily mean that dust isn't a prominent scatterer in the outer regions of these stars, just that the amount of dust does not account for the degree of polarisation seen. Geometrical considerations may be dominant, which might indicate that the stellar winds of later giants are less symmetrical, and consequently produce increasingly asymmetric dust-discs. In the model of \citet{johnson91} an asymmetric wind develops with a star's entry onto the asymptotic giant branch (AGB). Our observations here might indicate that it occurs sooner. Alternatively, a cooler atmosphere allows dust to form closer to the photosphere, which would produce a larger polarisation signal due to an increased scattering density.

\begin{figure}
\includegraphics[width=0.5\textwidth,natwidth=100,natheight=100,trim=0.5cm 0.5cm 0.5cm 0.5cm]{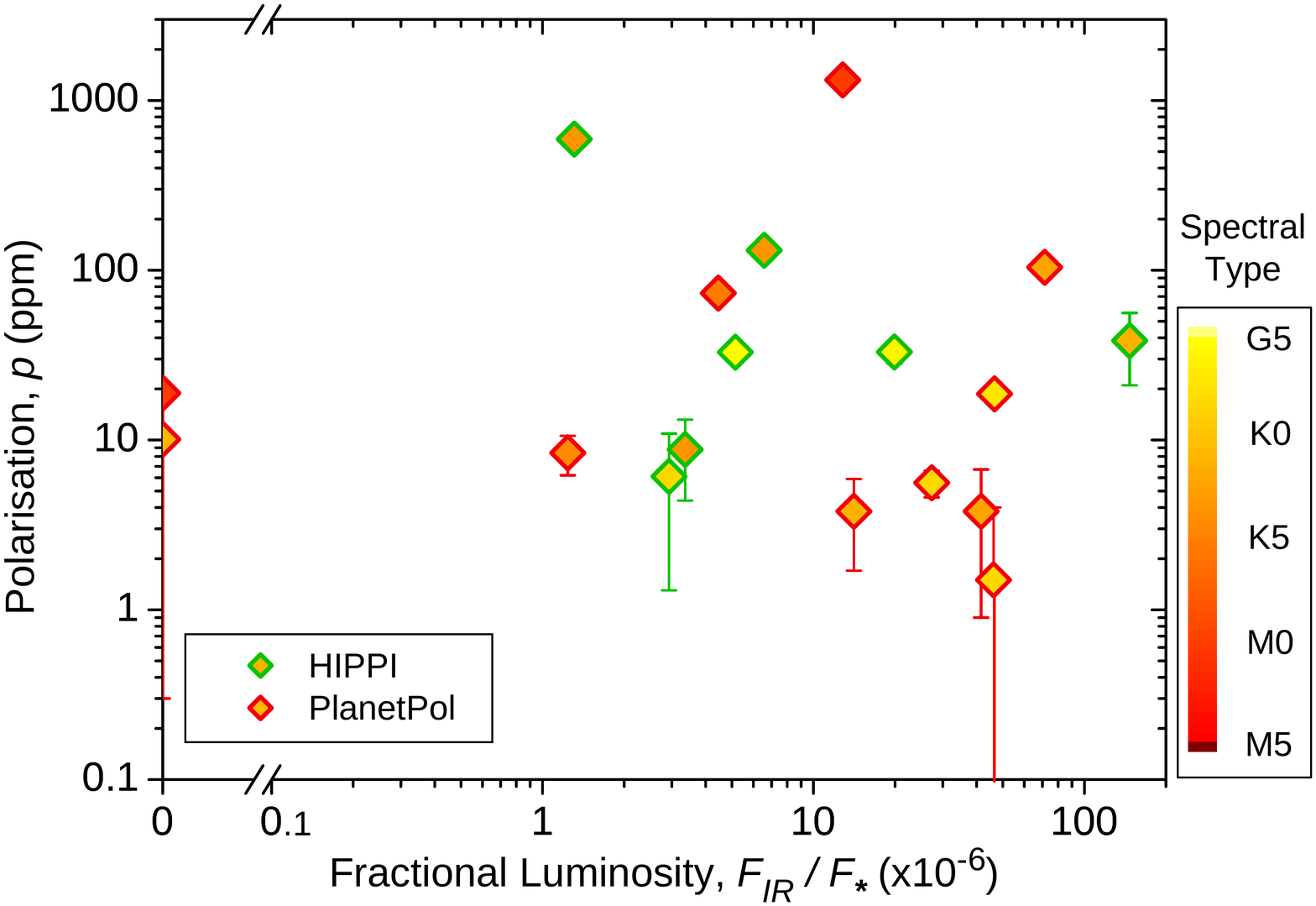}
\caption{Polarisation with infrared excess in giants later than G5. Note that HIPPI is 1.66 times as sensitive as PlanetPol at detecting polarisation caused by dust in red giants, but that it also sampled a region with a higher interstellar polarisation. Fractional luminosities come from \citet{mcdonald12}, other parameters for the PlanetPol stars come from \citet{bailey10}.}
\label{fig_FracLwP}
\end{figure}

\subsubsection{$\delta$ Oph and other mechanisms}

To investigate further we followed up our initial May 2014 observation of the suspected variable \citep{percy92} $\delta$ Oph with observations made in three different wavelength bands in May 2015. This object is particularly interesting because it features in the work of both \citet{tsuji08} on MOLspheres (extended molecular spheres thought to exist beyond the photosphere) and the work of \citet{ryde15} which largely refutes the need for a MOLsphere to explain spectral features such as metallic emission lines, OH lines and HF lines at 12 $\mu$m. \citet{ryde15} claimed that all of their sample, including $\delta$ Oph, were dust free, and used this as an argument against certain absorption features being produced in the MOLsphere against a continuum formed from an alumina dust shell. 

The measurements of $\delta$ Oph taken a year apart produced the same degree of polarisation within the error of the measurement as shown in Table \ref{tab_delta_oph}. However, the polarisation in \textit{q} was substantially different: -84.8$\pm$7.2 ppm for the later measurement compared to -15.4$\pm$6.0 ppm for the earlier one. The polarisation of $\delta$ Oph is therefore varying, even if the large, nearly static, value for \textit{u} masks the variation if just looking at the polarisation degree or angle. The follow up measurements reveal behaviour consistent with dust as a scatterer, i.e. an increase as $\lambda^{-1}$ between \textit{g$^{\rm \prime}$} and \textit{r$^{\rm \prime}$} before a less steep increase in polarisation for the 425SP filter; this is shown in Figure \ref{fig_DeltaOph}. The polarisation angle in all three bands is the same within the error of the measurements.

\begin{figure}
\includegraphics[width=0.5\textwidth,natwidth=100,natheight=100,trim=0.5cm 0.5cm 0.5cm 0.5cm]{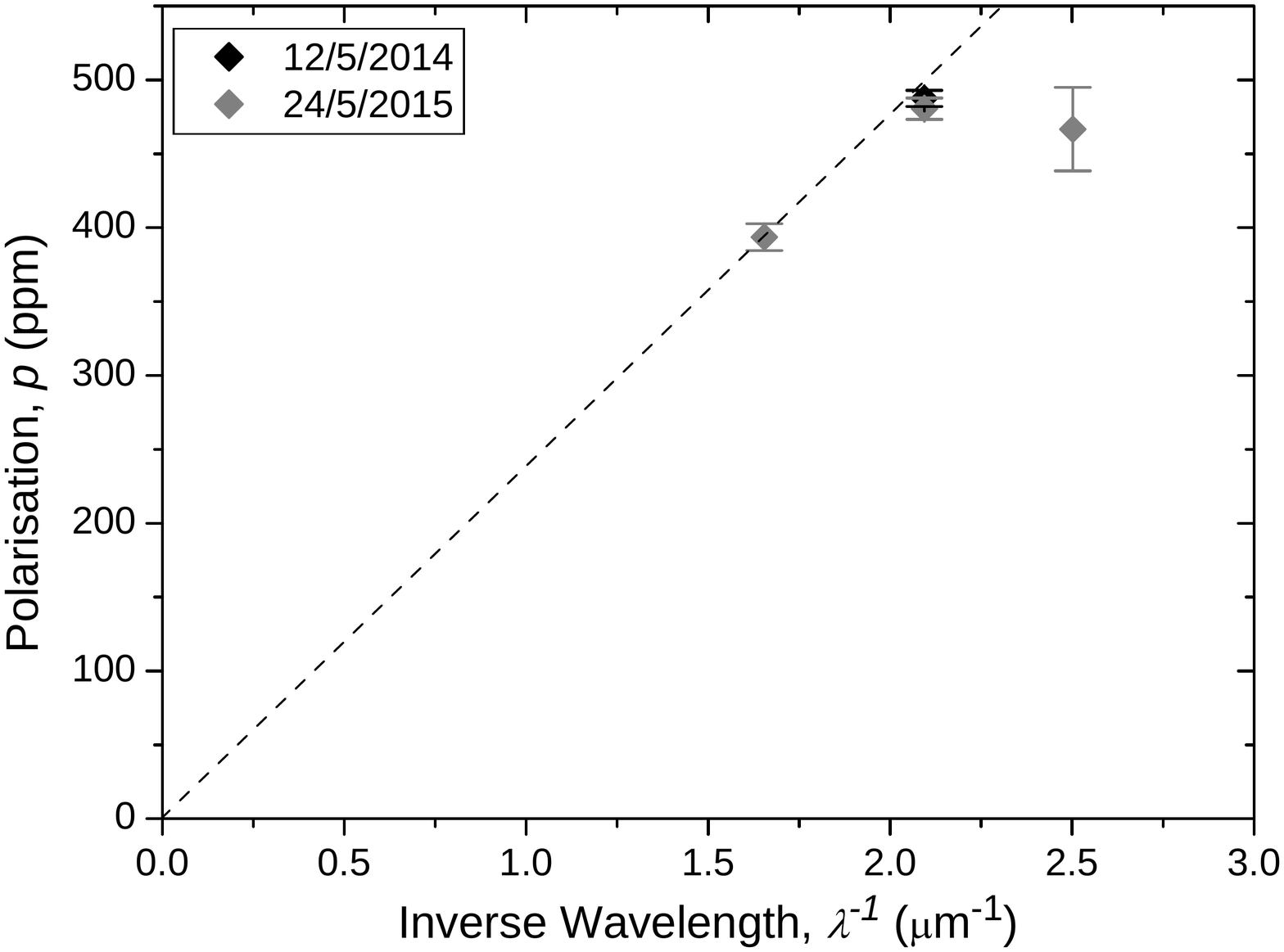}
\caption{polarisation measurements for $\delta$ Oph with inverse wavelength. The dashed guide-line is drawn from the origin through the first data point. The second data point lies close to the guide-line, showing a $\lambda^{-1}$ relationship.}
\label{fig_DeltaOph}
\end{figure}

\begin{table}
\caption{Individual measurements of $\delta$ Oph.}
\tabcolsep 2 pt
\centering
\begin{tabular}{lllrrrr}
\hline
Filter$^{a}$ & Date &
       \multicolumn{1}{c}{q (ppm)} & 
       \multicolumn{1}{c}{u (ppm)} & 
       \multicolumn{1}{c}{p (ppm)} &
       \multicolumn{1}{c}{$\theta$ ($\degr$)} \\
\hline
\textit{g$^{\rm \prime}$}  &	12/5/14   &   \hspace{1.5 mm}$-$15.4 $\pm$ \hspace{1.5 mm}6.0	&  $-$487.2 $\pm$ \hspace{1.5 mm}4.9	&  487.5 $\pm$ \hspace{1.5 mm}5.5	&	134.1 $\pm$ 0.7	\\
\hline	
\textit{r$^{\rm \prime}$}  &    24/5/15   &   \hspace{1.5 mm}$-$44.5 $\pm$ \hspace{1.5 mm}9.1   &  $-$391.2 $\pm$ \hspace{1.5 mm}9.2   &  393.7 $\pm$ \hspace{1.5 mm}9.1  &   131.8 $\pm$ 1.3 \\
\textit{g$^{\rm \prime}$}  &	24/5/15   &	\hspace{1.5 mm}$-$84.8 $\pm$ \hspace{1.5 mm}7.2	&  $-$473.0 $\pm$ \hspace{1.5 mm}7.1	&  480.6 $\pm$ \hspace{1.5 mm}7.2	&	129.9 $\pm$ 0.9	\\
425SP                      &    24/5/15   & $-$109.4 $\pm$ 28.3 & $-$433.1 $\pm$ 28.2 & 466.7 $\pm$ 28.2 &   127.9 $\pm$ 3.6 \\
\hline
\end{tabular}
\begin{flushleft}
a - The effective wavelength for each filter in $\mu$m$^{-1}$ is: 425SP: 2.502, \textit{g$^{\rm \prime}$}: 2.094,  \textit{r$^{\rm \prime}$}: 1.654. \\
\end{flushleft}
\label{tab_delta_oph}
\end{table}

\citet{clarke84,schwarz86} were able to show that circumstellar dust is not the only mechanism reponsible for polarisation of red giants. \citet{harrington69}, in reference to Mira variables, was the first to propose that Rayleigh scattering from the photosphere could be responsible for the observed polarisation. This mechanism was initially ruled out by \citet{dyck71} on the basis that the wavelength dependence was inconsistent with the $\lambda^{-4}$ dependence of Rayleight scattering. It wasn't until spectropolarimetry of supergiant Betelgeuse ($\alpha$ Ori) revealed polarisation that changed across, and was correlated with, spectral lines of, in particular, TiO \citep{clarke84,schwarz86}, and that the $\lambda^{-1}$ relationship seen was an artifact of \textit{broadband} polarimetry that this mechanism was accepted. This mechanism of Rayleigh or Thompson scattering from hot-spots or convection cells in the photosphere is able to explain polarisation that varies regularly (in regular variables) on time scales of months to years, whereas alternate mechanisms such as non-radial pulsations, or an equator to pole temperature gradient cannot \citep{schwarz86}. If there are multiple hot-spots a polarisation angle that varies fairly smoothly across bands can also be expected, however that is not the case here. Hot-spots as the origin of polarisation doesn't preclude the presence of additional scattering regions as well. In fact changes in polarisation across TiO lines point to TiO as an absorber or scatterer in the outer photosphere of highly polarised stars. It seems reasonable to expect that should hot-spots be the cause of polarisation in $\delta$ Oph, then molecules like water, be they present in an extended cool photosphere or in a MOLsphere would also have spectral lines with a polarisation different to the surrounding spectral continuum.

Starspots are also a possibility, with \citet{saar92} having calculated maximum polarisations for late giants from this mechanism of a few hundred ppm. Being evolved stars red giants are generally thought of as being inactive. However, a wavelet analysis carried out by \citet{hedges13} found evidence of starspots in 14 out of a subsample of 416 red giants observed by the Kepler satellite. Very recently significant magnetic activity has been detected in Mira \citep{vlemmings15}, and the RS CVn K giant V* XX Tri is famous for being observed with a `superspot' covering 11\% of its surface area \citep{kunstler15}. \citet{hedges13} analysis revealed features that could persist for a few hundred days. Even so, this is not the norm, and seems the most unlikely candidate for the polarisation seen in $\delta$ Oph. The consistency seen in our two \textit{g$^{\rm \prime}$} measurements a year apart would put $\delta$ Oph at the extreme end of the range with regard to the persistence of such features. The maximum polarisations calculated by \citet{saar92} rely on a filling factor of 24\%, and something approaching this level would have to be present to see a $\sim$500 ppm polarisation. This doesn't fit for a star not known to be magnetically active. 

Furthermore, similar degrees of polarisation are calculated for \textit{g$^{\rm \prime}$} by \citet{leroy90} due to differential saturation of spectral lines in magnetic fields. \citet{leroy90} was most interested in K dwarfs, but also looked at a K giant. That analysis shows that polarisation due to differential saturation should be greatest in the 425SP band -- which is not the case for $\delta$ Oph -- a worked example spectrum for Arcturus (BS 5340, K1.5III) with a hypothetical 500 Gauss field is shown. Arcturus was observed as part of the PlanetPol survey, recording 6.3$\pm$1.6 ppm; whereas in PlanetPol's waveband the calculation yields around 150 ppm. If this mechanism were primarily responsible for the polarisation we see in M and K giants we might expect the ratio of median polarisations determined by HIPPI and PlanetPol to be similar for M and K spectral types, which is not what we see in Table \ref{tab_spc}.

\section{Conclusions}

Fifty of the brightest stars in the southern hemisphere have been observed for their linear polarisation. We have investigated these alongside 49 northern stars from the PlanetPol survey. These two surveys are far more sensitive than previous stellar polarimetric surveys. Prior to this work, aside from the four classical Be stars, none of the stars had identified intrinsic polarisation assigned to them. We have now attributed polarigenic mechanisms to around half of the stars in the survey. In particular B type stars have been identified as being polarised through a variety of mechanisms, whilst late giants also show a greater propensity to be intrinsically polarised than demonstrated in previous studies, mostly as a consequence of circumstellar dust. Intrinsic polarisation was also identified in an Ap star and in debris disk systems.

Amongst the intrinsically polarised B stars was the close binary system Spica, which demonstrated variability from repeat measurements. Of the four Be stars observed, three have been resolved by interferometry. For each of these we measured a polarisation angle aligned perpendicular to the rotational axis. We recorded the highest degree of polarisation on record for the Be star Achernar in September of 2014; this indicates it was particularly active at this time. It has since declined in polarisation. Other B type stars were also found to possess high levels of polarisation. Since most of the B stars rotate at well below break-up velocity (Regulus in the PlanetPol survey being an exception) this cannot be a result of rotational oblateness. Instead we advance the hypothesis that what we are observing is low level Be-like behaviour in normal B stars.

There are only a small number of debris disk systems in the combined catalogue of the two surveys, but these show a slight tendency to be more polarised than other nearby stars. The circumbinary debris disk system $\epsilon$ Sgr appears to have a polarisation signature consistent with a companion-induced asymmetry in light scattered from the disk.

Many of the M and K giants in the surveys are clearly intrinsically polarised with cooler and more luminous stars having higher polarisations. Infrared excess is not correlated with the degree of polarisation in our sample. However circumstellar dust distributed by an asymmetric stellar wind may be responsible for the trends we see. As might an increased scattering density from dust closer to the photosphere in cooler stars. However the dust free M giant $\delta$ Oph also displays a variable and high degree of polarisation. This behaviour may be associated with Rayleigh scattering from stellar hot-spots, or less likely star spots.

The plethora of polarised stars identified has profound consequences for polarimetric studies of the local interstellar medium, and for searching out other polarimetric effects nearby. It is common practice to determine a value for the interstellar polarisation of an object thought to have some interesting intrinsic effect by taking an average of the measured polarisations of nearby stars. Stars are selected for this purpose on the basis that they are likely to be unpolarised. Our study shows that if one seeks a value for the interstellar polarisation at the ppm level then, in particular, no B type stars nor M or K giants should be selected. Ap stars and debris disk systems should also be avoided. As these are some of the brightest stars at any given distance and otherwise likely to be selected this is an important finding.

When we discount the intrinsically polarised stars from our sample and that of the PlanetPol survey, we find that interstellar polarisation is stronger in southern stars. Away from the Galactic equator in the northern sky stars are polarised at $\sim$2 $\times$10$^{-7}$ pc$^{-1}$. At distances between 10 and 30 pc in the southern sky the rate of polarisation can be as much as $\sim$2 $\times$10$^{-6}$ pc$^{-1}$. Beyond 30 pc there appears to be less of a polarising effect from the interstellar medium. The polarisation angles of stars just to the Galactic north are aligned to the Galactic equator. A dustier interstellar medium to the south (in the direction of the centre of the galactic plane) appears primarily responsible for the higher degrees of interstellar polarisation observed in southern stars within 100 pc.

The very high precision of HIPPI in fractional polarisation has allowed us to uncover the polarimetric behaviour of some of the best known stars in the night sky. We have demonstrated HIPPI's potential in many areas of stellar polarimetry, producing more precise measurements than have previously been possible and indeed uncovering the polarigenic behaviour of some stars previously searched for but not found. HIPPI's simplicity and economical fabrication belies its great value.

\section*{Acknowledgments}

The development of HIPPI was funded by the Australian Research Council through Discovery Projects  grant DP140100121 and by the UNSW Faculty of Science through its Faculty Research Grants program. The authors thank the Director and staff of the Australian Astronomical Observatory for their advice and support with interfacing HIPPI to the AAT and during our observing runs on the telescope.  We acknowledge the use of the SIMBAD database and this research has made use of the VizieR Service at Centre de Donn\'{e}es Astronomiques de Strasbourg. 

\bibliographystyle{mnras}
\bibliography{hippi_bright_star}

\bsp

\label{lastpage}

\end{document}